\documentclass[lettersize,journal]{IEEEtran}
\usepackage{amsmath,amsfonts}
\usepackage{relsize}
\usepackage{algorithmic}
\usepackage{algorithm}
\usepackage{array}
\usepackage[caption=false,font=normalsize,labelfont=sf,textfont=sf]{subfig}
\usepackage{textcomp}
\usepackage{stfloats}
\usepackage{url}
\usepackage{verbatim}
\usepackage{graphicx}
\usepackage{cite}
\usepackage{booktabs}
\usepackage{multirow}
\usepackage{threeparttable}

\usepackage{amssymb}
\usepackage{float}
\usepackage{listings}
\usepackage{tcolorbox}
\tcbuselibrary{listingsutf8}

\usepackage{xcolor}
\usepackage[table]{xcolor} 
\definecolor{OursColor}{HTML}{E0EFFF} 
\definecolor{SotaColor}{HTML}{FFF8E1} 
\begin{document}

\title{Therm-FM: A Foundation-Model Framework for Unified Steady-State and Transient Thermal Simulation in 3D-ICs}

\title{Therm-FM: PDE Foundation Model Is All You Need for \\ 
3D-ICs Thermal Simulation}

\title{Therm-FM: PDE Foundation Model Adaptation for Data-Efficient 3D-IC Thermal Simulation}

\title{Therm-FM: Data-Efficient 3D-IC Thermal Simulation via PDE Foundation Model Adaptation}

\title{Therm-FM: Foundation Model is ALL YOU NEED \\ for 3D-ICs Thermal Simulation}


\author{
Zhen Huang,
Haiyang Xin,
Wenkai Yang,
Yangbo Wei,
Zhiping Yu,~\IEEEmembership{Life Fellow,~IEEE}, \\
Yu Zhang,
Wei W. Xing,
Ting-Jung Lin,
and Lei He,~\IEEEmembership{Fellow,~IEEE}%
\thanks{
This work is an extended version of our preliminary conference paper~\cite{huang2026pde} which has been accepted by DAC 2026.
}
\thanks{
Zhen Huang, Haiyang Xin and Yu Zhang are with the School of Computer Science and Technology,
University of Science and Technology of China, Hefei, China.
}%
\thanks{
Zhen Huang, Wenkai Yang, Yangbo Wei and Lei He are with the Eastern Institute of Technology,
Ningbo, China.}

\thanks{
Zhiping Yu is with Tsinghua University, Beijing, China.
}%
\thanks{
Wei W. Xing is with The University of Sheffield, Sheffield, United Kingdom.
}%
\thanks{
Ting-Jung Lin is with Ningbo Institute of Digital Twin, Eastern Institute of Technology,
Ningbo, China, and also with the Engineering Research Center of Chiplet Design and
Manufacturing of Zhejiang Province, Ningbo, China.
}%

\thanks{
Corresponding authors:  Ting-Jung Lin and Wei W. Xing 
}%

}

\maketitle
\begin{abstract}

Data-driven thermal predictors for 3D-ICs are often trained from scratch for each chip design using many high-fidelity finite-element simulations, leading to high data-generation cost and expensive cross-design reuse. We propose Therm-FM, a neural operator framework that adapts a pretrained partial differential equation(PDE) foundation model to steady-state and transient 3D-IC thermal simulation. The motivation is that chip-level heat conduction shares elliptic and parabolic operator structures with diffusion-type PDEs, allowing pretrained diffusion priors to initialize thermal-field prediction despite heterogeneous materials, dense TSV/\textmu{}bump interconnects, and package-level boundary conditions. To further reduce data-generation cost, Therm-FM incorporates a thermal-equivalent multi-fidelity training strategy that uses low-cost approximate simulations for thermal-domain adaptation and limited high-fidelity samples for calibration. Experiments on public HotSpot benchmarks and industrial 3D-IC package benchmarks show that Therm-FM achieves up to a $10.6\times$ reduction in mean error and surpasses prior best accuracy with less than 20\% of the training data. In cross-chip adaptation, it matches or surpasses full-data baselines in several metrics using only 10--30 target samples. We release datasets, source code, and pretrained models at \url{https://github.com/haiyangxin/Therm-FM}.

\end{abstract}


\section{Introduction}

Thermal simulation is a repeated and costly step in 3D-IC design workflows.
Traditional finite element method (FEM) solvers provide accurate solutions but require minutes to hours per evaluation~\cite{FEM1,FEM2}, while practical design-space exploration often requires hundreds to thousands of evaluations~\cite{floorplan1}.

\begin{figure}[t]
    \centering
    \includegraphics[width=0.812\linewidth]{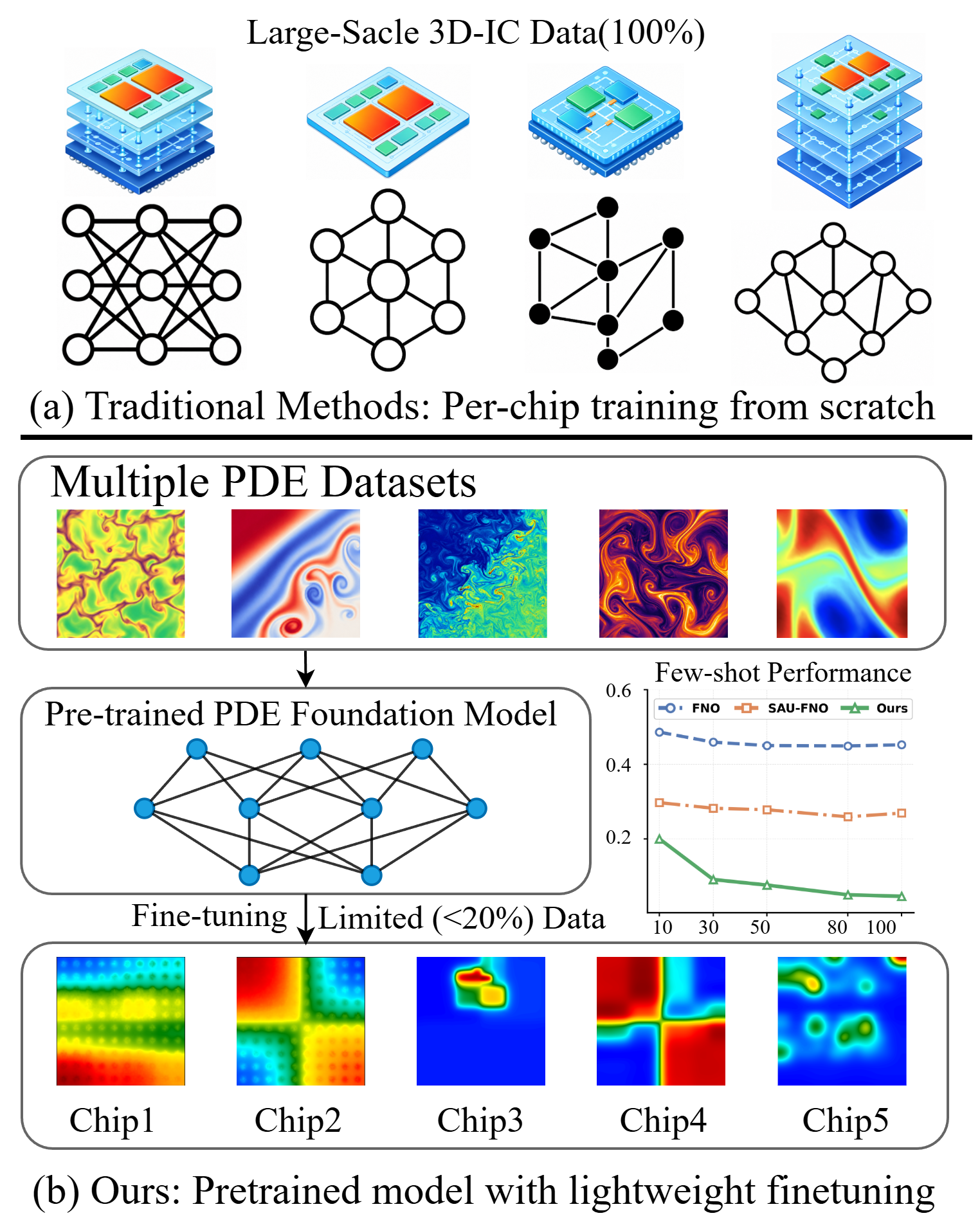}
    \caption{
        From traditional per-chip training from scratch to Therm-FM: fine-tuning a pretrained PDE foundation model for 3D-IC thermal simulation.
    }
    \vspace*{-8pt}
    \label{fig:intro}
\end{figure}

Learning-based thermal surrogates, ranging from CNN-based predictors and physics-constrained models to neural operators, can substantially reduce inference time once trained~\cite{deepoheat,saufno}.
However, despite architectural progress, these methods remain largely design-specific: changes in chip structures, material properties, or operating conditions typically require new high-fidelity FEM data and retraining from scratch, as illustrated in Fig.~\ref{fig:intro}(a).
This per-design data and training cost reflects a structural limitation of current learning-based thermal simulation pipelines~\cite{FEM3,ML-2}.


A structural observation motivates a different approach. 
The steady-state heat conduction equation in 3D-ICs,
\begin{equation}
\nabla \cdot (k(x)\nabla T(x)) + Q(x) = 0,
\end{equation}
shares the same elliptic operator form as the general diffusion equation
\begin{equation}
\nabla \cdot (\kappa(x)\nabla u(x)) + f(x) = 0,
\end{equation}
under the correspondence $k \leftrightarrow \kappa$, $T \leftrightarrow u$, and $Q \leftrightarrow f$. 
The transient case follows analogously. 
This is not a loose analogy but a structural equivalence at the operator level. It implies that spatial coupling patterns, source–response structures, and temporal propagation behaviors learned from diffusion-type PDEs during pretraining transfer directly to chip-level thermal fields. 
Analogously to how LLMs in natural language processing are pretrained on large text corpora and then fine-tuned with modest labeled data for specific downstream tasks, PDE foundation models~\cite{hao2024dpot,herde2024poseidon} are pretrained on diverse PDE datasets to learn transferable physical priors, enabling adaptation to downstream physics problems with limited target data. Yet, whether this capability extends to 3D-IC thermal simulation remains unexplored. If confirmed, the practical impact would be substantial: a single pretrained model could be fine-tuned with limited target data for each new chip design, reducing the need for per-chip training from scratch.

To address these challenges, we propose \textbf{Therm-FM}, a framework that combines a neural operator backbone, which maps input physical fields to output temperature fields without assuming a fixed spatial resolution, with a pretrained PDE foundation model to enable data-efficient 3D-IC thermal simulation across diverse chip architectures as illustrated in Fig.~\ref{fig:intro}(b). By exploiting diffusion-oriented physical priors acquired during pretraining, Therm-FM eliminates the need for per-chip training from scratch, requiring only limited target data for fine-tuning, and unifies steady-state and transient analysis within a single formulation.
To reduce the cost of generating high-fidelity simulation data, we further devise a multi-fidelity training strategy: an analytical thermal-equivalent model produces inexpensive low-fidelity data, which are fused with sparse high-fidelity simulations to enhance target-domain adaptation.

We evaluate Therm-FM on public HotSpot benchmarks and industrial 3D-IC package cases across different structures, resolutions, and steady-state/transient tasks.
Therm-FM consistently outperforms existing thermal predictors, achieving up to a $10.6\times$ reduction in mean error, surpassing SOTA accuracy with less than 20\% of the training data, and showing strong cross-chip adaptation with only 10--30 target samples.

The main contributions of this work are summarized as follows.

\begin{itemize}

\item \textbf{PDE foundation-model adaptation for 3D-IC thermal simulation.}
We identify the operator-level alignment between 3D-IC heat conduction and diffusion-type PDEs, and adapt a pretrained PDE foundation model to steady-state and transient thermal prediction.
Therm-FM surpasses SOTA with fewer than 20\% of the training samples and up to a $10.6\times$ reduction in mean error.

\item \textbf{Thermal-equivalent modeling and multi-fidelity learning.}
We develop a thermal-equivalent model and integrate it into a multi-fidelity learning strategy, enabling efficient adaptation with reduced high-fidelity FEM data.

\item \textbf{Comprehensive evaluation and open release.}
We evaluate Therm-FM on public HotSpot benchmarks and newly constructed industrial 3D-IC package cases across different structures, resolutions, and steady-state/transient tasks. Results show consistent accuracy gains, strong cross-chip adaptation with only 10--30 target samples, and favorable scaling behavior.

\end{itemize}

\section{Related Work}

\subsection{Machine Learning for Thermal Simulation}

Machine learning methods typically formulate thermal simulation as a supervised regression problem from input design to temperature fields~\cite{chandra20252d,coenen2025pindas}.
Early CNN- and encoder--decoder-based models enable fast inference, but often rely on fixed layouts, fixed grids, and specific training distributions~\cite{CNNthermal,FNO1,wang2024transfer}.
When chip structures, materials, or boundary conditions change, these models usually require training from scratch for the target design.

PINNs and physics-constrained models incorporate heat-conduction equations into training to improve physical consistency~\cite{PINN23,yang2026physics,thermal-pinn}.
However, multi-layer 3D-ICs introduce heterogeneous materials, dense interconnects, and multi-scale thermal coupling, which lead to discontinuous PDE residuals, imbalanced losses across conductivity regions, and slow optimizer convergence.
Consequently, PINN-based methods often suffer from unstable training and high computational cost~\cite{PINN1,pinn-2,2025-pinn-dac}.


\subsection{Neural Operators and PDE Foundation Models}

Neural operators learn mappings from input functions to solution functions, providing a general framework for PDE-governed systems.
Representative methods such as FNOs~\cite{FNO}, DeepONets~\cite{lu2019deeponet}, and their thermal-field variants~\cite{ML-3,ML-5,thermal-transformer,thermal25-compactmodel,thermal-2025fast,thermal24-chip} have improved prediction accuracy and cross-resolution modeling in scientific and thermal simulation tasks.
However, conventional neural operators are typically trained from random initialization for each target problem, making them dependent on large amounts of high-fidelity data and limiting their cross-design transfer.

Recent PDE foundation models extend neural operators beyond training from scratch through a pretraining--fine-tuning paradigm~\cite{hao2024dpot,herde2024poseidon}.
By learning transferable physical priors from diverse PDE datasets, these models can adapt to downstream PDE tasks with limited target data.
However, their effectiveness in 3D-IC thermal simulation remains underexplored, especially under heterogeneous materials, dense TSV/\textmu{}bump interconnects, multi-layer stacked package structures, and engineering-level boundary conditions.
This work investigates this gap by adapting a pretrained PDE foundation model to 3D-IC thermal simulation and combining it with thermal-equivalent modeling and multi-fidelity training for data-efficient cross-chip thermal prediction.

\section{Therm-FM Framework}

\begin{figure*}[htbp]
  \centering
  \includegraphics[width=\textwidth]{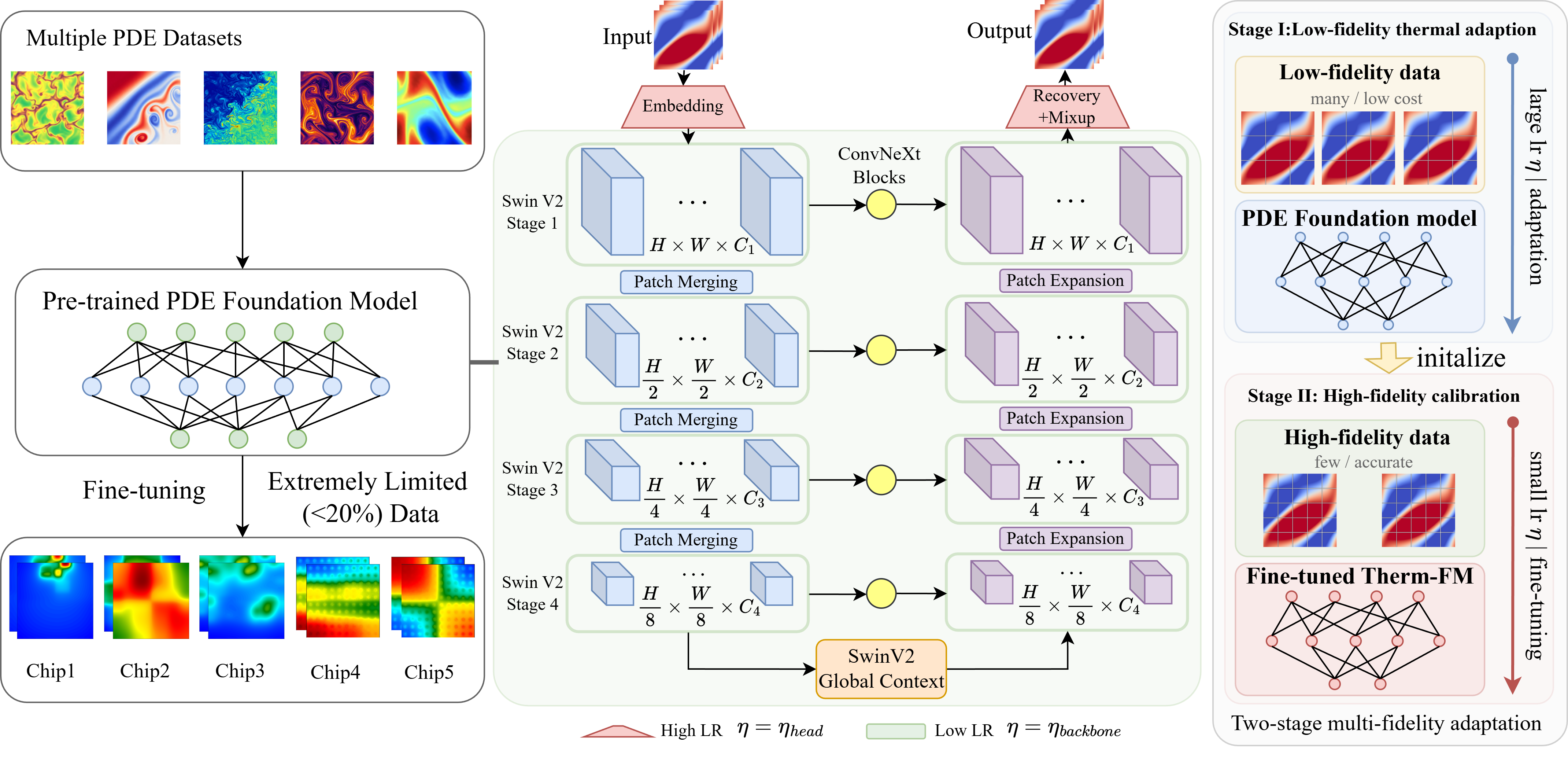}
\caption{
Workflow of Therm-FM. 
The left panel shows PDE foundation-model pretraining and lightweight fine-tuning for 3D-IC thermal prediction.
The middle panel illustrates the neural-operator backbone with embedding/recovery layers and hierarchical SwinV2 blocks.
The right panel presents the two-stage multi-fidelity adaptation process using low- and high-fidelity data.
}
  \label{fig:Therm-FM}
\end{figure*}

We formulate thermal simulation as an operator learning problem that predicts temperature fields from power distributions, material/structural information, and optional temporal conditions. 
Therm-FM is built upon a pretrained PDE foundation model and adapts it to chip-level thermal analysis using limited target-domain thermal data.

\subsection{Problem Setup}

The thermal behavior in 3D-ICs is governed by heat conduction. 
Let $\Omega \subset \mathbb{R}^3$ denote the spatial domain and $x=(x,y,z)$ denote the spatial coordinates. 
The transient temperature field $T(x,t)$ satisfies
\begin{equation}
\rho C_p \frac{\partial T(x,t)}{\partial t} =
\nabla \cdot (k(x)\nabla T(x,t)) + Q(x,t),
\end{equation}
where $\rho$, $C_p$, and $k(x)$ denote material density, heat capacity, and spatially varying thermal conductivity, respectively, and $Q(x,t)$ is the power density.

For steady-state analysis, the temporal derivative vanishes, and the governing equation reduces to
\begin{equation}
\nabla \cdot (k(x)\nabla T(x)) + Q(x) = 0.
\end{equation}

From an operator learning perspective, our goal is to learn a mapping from input physical fields to the corresponding temperature field. 
Specifically, the input includes the power distribution $Q$, material or structural information $M$, and an optional temporal condition $t$, while the output is the temperature field $T$:
\begin{equation}
G: (Q,t;M) \rightarrow T ,
\end{equation}
In our benchmarks, $M$ is fixed within each case and is implicitly captured through adaptation rather than provided as an explicit input channel. The steady-state task can be regarded as a special case without the temporal dimension, whereas the transient task learns the mapping from power-time inputs to temperature-time responses. 


\subsection{Framework Overview}

The overall workflow of Therm-FM is illustrated in Fig.~\ref{fig:Therm-FM}. 
Given the power distribution and associated physical inputs, the model first embeds them into a latent representation. 
The latent features are then processed by a neural operator backbone initialized from a pretrained PDE foundation model, which captures global spatial dependencies and local thermal couplings induced by heat diffusion. 
Finally, a recovery layer projects the latent features back to the physical space to obtain the predicted temperature field.

The key idea of Therm-FM is to transfer diffusion-oriented physical priors learned during PDE pretraining to chip-level heat conduction. 
For a new chip design, the model is not trained from random initialization; instead, it is fine-tuned from the pretrained backbone using a limited number of target-domain samples. 
For transient thermal simulation, we incorporate the temporal dimension into the input and output representations, allowing the model to predict temperature evolution over multiple time steps. 
This extension keeps the backbone architecture unchanged and only adjusts task-specific input and output mappings. 
As a result, Therm-FM provides a unified framework for both steady-state and transient thermal simulation, adapting to new chip designs with only limited target-domain samples.

\subsection{Foundation Model Adaptation}



Therm-FM adapts Poseidon~\cite{herde2024poseidon}, a pretrained PDE foundation model trained on large-scale fluid-dynamics PDE data, to 3D-IC thermal simulation via lightweight fine-tuning.
Its learned spatiotemporal operator priors are transferable to chip-level heat conduction due to the shared diffusion-type operator structure.
A general diffusion equation with a spatially varying capacity coefficient and source term can be written as
\begin{equation}
c(x)\frac{\partial u(x,t)}{\partial t}
=
\nabla \cdot \bigl(\boldsymbol{\kappa}(x)\nabla u(x,t)\bigr) + f(x,t),
\end{equation}
where $u$ is the state variable, $c(x)$ represents the spatially varying storage capacity, $\boldsymbol{\kappa}(x)$ is the spatially varying diffusion tensor, and $f(x,t)$ is the external source term. 
The transient heat-conduction equation in heterogeneous 3D-ICs can be formulated within the same conservative diffusion-type structure under the following correspondence:
\begin{equation}
\begin{aligned}
    u &\leftrightarrow T, \qquad & c(x) &\leftrightarrow \rho(x) C_p(x), \\
    \boldsymbol{\kappa}(x) &\leftrightarrow \mathbf{k}(x), \qquad & f(x,t) &\leftrightarrow Q(x,t).
\end{aligned}
\end{equation}
Here, $\mathbf{k}(x) = \mathrm{diag}(k_x(x), k_y(x), k_z(x))$ is the anisotropic thermal conductivity tensor, which naturally accommodates the direction-dependent effective conductivities derived from our homogenization strategy. 
Therefore, diffusion-related priors learned during PDE pretraining, such as spatial coupling, anisotropic diffusion responses, and temporal propagation patterns, can benefit transient thermal prediction.

For steady-state analysis, the temporal derivative vanishes, and the general diffusion equation reduces to
\begin{equation}
\nabla \cdot \bigl(\kappa(x)\nabla u(x)\bigr) + f(x) = 0,
\end{equation}
which shares the same elliptic operator form as the steady-state heat-conduction equation
\begin{equation}
\nabla \cdot \bigl(k(x)\nabla T(x)\bigr) + Q(x) = 0.
\end{equation}
This alignment further supports the transfer of pretrained priors related to steady-state field structures, source-induced responses, and diffusion under material heterogeneity.

Let $B_{\theta_0}$ denote the neural operator backbone pretrained on a source PDE distribution $\mathcal{D}_{\mathrm{source}}$. 
Because the pretraining tasks and 3D-IC thermal simulation differ in input channels, material/structural information, output resolution, and boundary conditions, Therm-FM decomposes the adapted model into an input embedding layer, a pretrained backbone, and an output recovery layer:
\begin{equation}
\hat{T}
=
\phi_{\mathrm{rec}}
\circ B_{\theta}
\circ \phi_{\mathrm{emb}}(Q,t;M),
\end{equation}
where $Q$ denotes the power distribution, $t$ denotes an optional temporal condition, and $M$ denotes the fixed chip/package configuration.
For steady-state prediction, the temporal condition is omitted; for transient prediction, the temporal condition or temporal dimension is incorporated to predict temperature responses over multiple time steps.

During adaptation, $\phi_{\mathrm{emb}}$ and $\phi_{\mathrm{rec}}$ are reinitialized to match the input and output format of the 3D-IC thermal task, while $B_{\theta}$ is initialized from the pretrained weights $B_{\theta_0}$ and further fine-tuned on the target data. 
Given a target thermal distribution $\mathcal{D}_{\mathrm{target}}$, the adaptation objective is
\begin{equation}
\theta^{*}, \phi_{\mathrm{emb}}^{*}, \phi_{\mathrm{rec}}^{*}
=
\arg\min_{\theta,\phi_{\mathrm{emb}},\phi_{\mathrm{rec}}}
\mathbb{E}_{(Q,M,t,T)\sim \mathcal{D}_{\mathrm{target}}}
\left[
\mathcal{L}\bigl(\hat{T},T\bigr)
\right],
\end{equation}
Compared with training from random initialization, this adaptation process preserves the diffusion-oriented physical prior learned during pretraining and aligns it with the target thermal data distribution using limited samples. 
The same formulation applies to both steady-state and transient thermal tasks, and also provides a unified interface for incorporating high-fidelity samples and low-fidelity samples generated by the thermal-equivalent model.

\subsection{Thermal-Equivalent Modeling}
\label{sec:strategy}

\begin{figure*}[htbp]
  \centering
  \includegraphics[width=\textwidth]{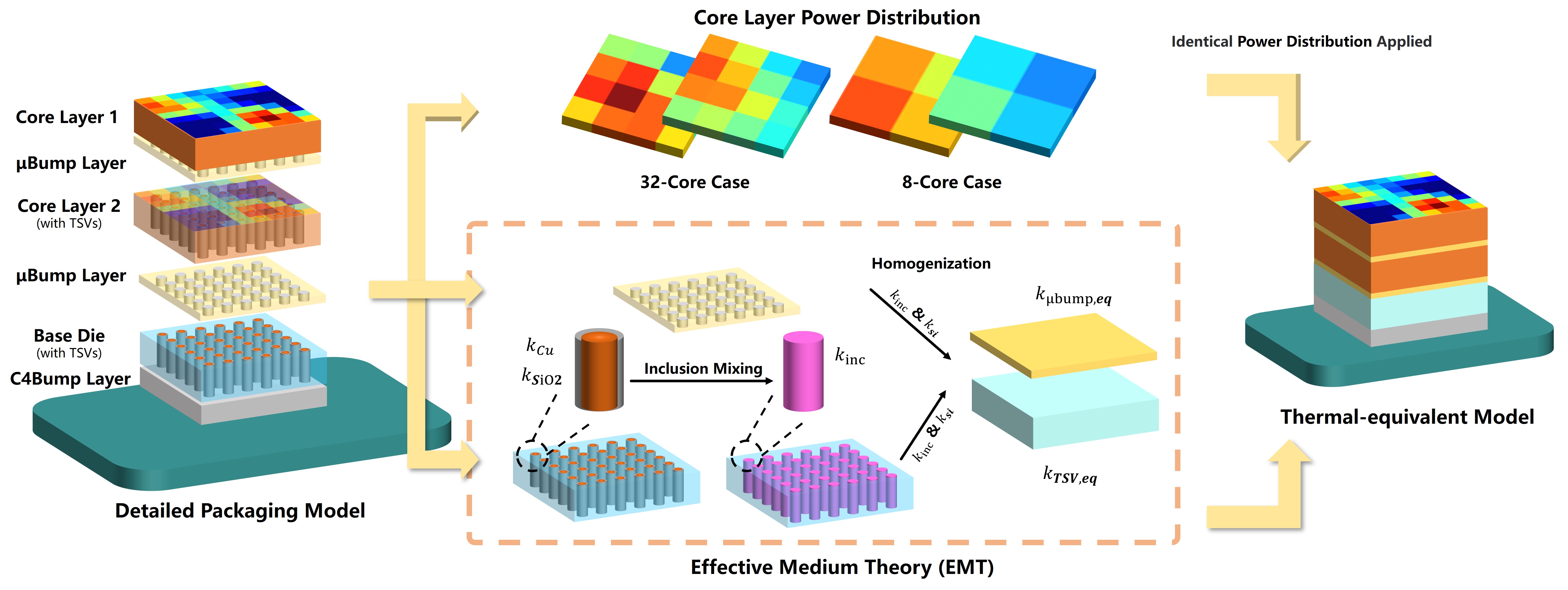}
\caption{
Industrial 3D-IC benchmark construction and EMT-based low-fidelity modeling.
The detailed package contains heterogeneous core, $\mu$bump, TSV, C4 bump, and substrate layers.
Core-level power distributions are applied to both the detailed and thermal-equivalent models, where dense TSV and $\mu$bump arrays are homogenized into macroscopic equivalent layers for low-cost thermal simulation.
}
  \label{fig:newcase}
  \vspace*{-10pt}

\end{figure*}

To reduce the cost of high-fidelity thermal data generation, we introduce an analytical thermal-equivalent model for multi-fidelity training.
As shown in Fig.~\ref{fig:newcase}, dense TSV and $\mu$bump arrays introduce micrometer-scale multi-material structures, requiring very fine meshes in detailed FEM models.
We therefore use effective medium theory (EMT)~\cite{10.1093/acprof:oso/9780198705093.001.0001} to homogenize these interconnect arrays into macroscopic anisotropic layers.
The resulting equivalent model preserves the same power distributions and package-level boundary conditions as the detailed high-fidelity model, while substantially reducing the mesh complexity.

For vertical conduction, heat flux is approximately parallel to the interconnect pillars, so the effective conductivity is computed by a mixing rule:
\begin{equation}
    k_{z,\mathrm{eq}} =
    f_{\mathrm{Cu}} k_{\mathrm{Cu}}
    + f_{\mathrm{ox}} k_{\mathrm{ox}}
    + f_{\mathrm{Si}} k_{\mathrm{Si}},
\end{equation}
where $f_{\mathrm{Cu}}$, $f_{\mathrm{ox}}$, and $f_{\mathrm{Si}}$ are the volume fractions of the copper core, SiO$_2$ shell, and silicon matrix with respect to the entire macroscopic layer.

For lateral conduction, cylindrical TSVs bend heat-flow streamlines, so we use a two-step EMT approximation.
First, the core--shell TSV is converted into an equivalent inclusion~\cite{hiroshi1986equivalent}:
\begin{equation}
    k_{\mathrm{inc}} =
    k_{\mathrm{ox}}
    \left[
    \frac{
    k_{\mathrm{Cu}} + k_{\mathrm{ox}} + v_{\mathrm{core}}(k_{\mathrm{Cu}} - k_{\mathrm{ox}})
    }{
    k_{\mathrm{Cu}} + k_{\mathrm{ox}} - v_{\mathrm{core}}(k_{\mathrm{Cu}} - k_{\mathrm{ox}})
    }
    \right],
\end{equation}
where $v_{\mathrm{core}}=(r_{\mathrm{Cu}}/r_{\mathrm{ox}})^2$ and $r_{\mathrm{ox}}$ is the outer radius of the SiO$_2$ shell.
The equivalent inclusions are then embedded into the silicon matrix using the 2-D Maxwell--Eucken model~\cite{gong2014novel}:
\begin{equation}
    k_{x,\mathrm{eq}} = k_{y,\mathrm{eq}} =
    k_{\mathrm{Si}}
    \left[
    \frac{
    k_{\mathrm{inc}} + k_{\mathrm{Si}} + f_{\mathrm{TSV}}(k_{\mathrm{inc}} - k_{\mathrm{Si}})
    }{
    k_{\mathrm{inc}} + k_{\mathrm{Si}} - f_{\mathrm{TSV}}(k_{\mathrm{inc}} - k_{\mathrm{Si}})
    }
    \right],
\end{equation}
where $f_{\mathrm{TSV}}=f_{\mathrm{Cu}}+f_{\mathrm{ox}}$.
For $\mu$bump layers, the solder bumps are modeled as vertical cylindrical inclusions embedded in an underfill matrix. We apply the identical parallel mixing rule and Maxwell--Eucken model for their vertical and lateral homogenization, respectively. 
Furthermore, microscopic interfacial thermal resistance is not explicitly resolved in the analytical EMT model; its impact is partially reflected through the validated effective response, and residual bias is corrected by the high-fidelity calibration stage.

For transient simulation, the effective volumetric heat capacity is computed by volume averaging:
\begin{equation}
    C_{v,\mathrm{eq}} =
    \sum_m f_m C_{v,m},
\end{equation}
where $f_m$ and $C_{v,m}$ denote the volume fraction and volumetric heat capacity of material $m$.

With these equivalent parameters, TSV and $\mu$bump layers are replaced by macroscopic thermal layers, enabling fast generation of low-fidelity steady-state and transient data.
These low-fidelity samples capture large-scale diffusion patterns and are later combined with limited high-fidelity FEM samples for fine-grained calibration.
This overall EMT approximation assumes regular array distributions, scale separation, and locally-averaged thermal equilibrium, and its accuracy is validated in Sec. IV-D.

\subsection{Multi-fidelity Training}

To reduce the dependence on expensive high-fidelity simulations, we adopt a two-stage multi-fidelity training strategy that first adapts Therm-FM with low-fidelity data generated by the thermal-equivalent model and then calibrates it with limited high-fidelity FEM samples.

In the first stage, Therm-FM is initialized from the pretrained PDE foundation model and adapted using low-fidelity data. 
These temperature fields, which are generated by the analytical thermal-equivalent model at a fraction of the FEM cost, capture the dominant heat-diffusion trends and global temperature distributions.
This stage helps bridge the gap between the general PDE pretraining distribution and the target 3D-IC thermal simulation domain, including chip-level power inputs, packaging structures, and thermal-field outputs.

In the second stage, the model is further fine-tuned using a small set of high-fidelity data. 
Compared with low-fidelity data, high-fidelity simulations more accurately capture material heterogeneity, geometric details, local interconnect effects, and boundary conditions. 
This stage calibrates the bias introduced by the equivalent model and improves prediction accuracy on realistic high-fidelity thermal fields.
To stabilize calibration under limited high-fidelity supervision, we use a reduced learning rate in this stage, preventing large parameter updates while refining high-fidelity thermal details.

Let $\mathcal{D}_l^M$ and $\mathcal{D}_h^M$ denote the low- and high-fidelity thermal data distributions for a fixed chip/package configuration $M$, respectively. Both stages use the same supervised prediction objective:
\begin{equation}
\mathcal{L}_{\mathcal{D}^{M}}(\theta)
=
\mathbb{E}_{(Q,t,T)\sim \mathcal{D}^{M}}
\left[
\left\|G_{\theta}(Q,t;M)-T\right\|_2^2
\right],
\label{eq:mf_objective}
\end{equation}
where $\mathcal{D}^M=\mathcal{D}_l^M$ in the first stage and $\mathcal{D}^M=\mathcal{D}_h^M$ in the second stage.
For steady-state prediction, the temporal condition $t$ is omitted; for transient prediction, it is incorporated to supervise temperature responses over multiple time steps.


In this way, low-fidelity data provide coarse thermal-domain adaptation, while limited high-fidelity samples calibrate fine-grained thermal responses, reducing data-generation cost without sacrificing accuracy.

\section{Experiments}

\subsection{Dataset and Benchmark Configurations}
\label{sec:case}

We evaluate Therm-FM on two representative benchmark suites: public HotSpot benchmarks commonly used in prior work~\cite{wang2024aro, saufno} for fair comparison, and newly constructed industrial 3D-IC package benchmarks for assessing its generality and robustness on more complex multi-layer stacked structures with dense TSV/\textmu{}bump interconnects.
For clarity, \textbf{HS-SC}, \textbf{HS-QC}, and \textbf{HS-OC} denote the HotSpot single-core, quad-core, and octa-core cases, respectively; \textbf{IND-8C} and \textbf{IND-32C} denote the industrial 8-core and 32-core cases.

\textbf{HotSpot Benchmarks.}
We adopt 3D-IC thermal benchmarks generated by the open-source HotSpot simulator~\cite{huang2006hotspot}, a compact thermal resistance/capacitance network-based tool widely used in learning-based thermal simulation.
The floorplans are derived from the Alpha 21264 EV6 processor~\cite{kessler1999alpha64} and include single-core, quad-core, and octa-core stacking configurations, covering different design scales and thermal coupling complexities.
The models include thermal interface material (TIM) and TSV structures, with adiabatic lateral boundaries and convective cooling at the top surface.
Following prior settings, each steady-state case contains 5000 samples, split into 3600/400/1000 samples for training, validation, and testing.
We also evaluate transient thermal prediction on the corresponding HotSpot configurations, where each sample contains a temporal temperature sequence under time-varying power inputs.
Each benchmark setting uses the spatial and temporal resolutions adopted in prior work to ensure fair comparison.
Resolutions reported throughout, such as $55\times55$ or $88\times88\times9$, denote the output grid size and, for transient tasks, the number of temporal frames.

\textbf{Industrial 3D-IC Benchmarks.}
To evaluate Therm-FM under more realistic packaging scenarios, we construct two industrial 3D-IC package cases, IND-8C and IND-32C. 
As shown in Fig.~\ref{fig:newcase}, both cases adopt a stacked package structure consisting of two active core layers, one base die, $\mu$bump layers, TSVs with SiO$_2$ isolation, a C4 bump layer, and a substrate. 
The geometric parameters are summarized in Table~\ref{tab:new_benchmarks}. 
For IND-8C and IND-32C, each active layer is partitioned into 4 and 16 cores, respectively, corresponding to 8 and 32 total cores. 
To simulate non-uniform workloads, the power of each core is randomly sampled from $[0, 0.04]\ \mathrm{W}$ and applied to the corresponding core region as a volumetric heat source.

\begin{table}[tbp]
\smaller
  \centering
  \caption{Geometric parameters of the industrial 3D-IC benchmarks.}
  \label{tab:new_benchmarks}
  \begin{tabular}{lccc}
    \toprule
    \textbf{Layer} & \textbf{Area (mm$^2$)} & \textbf{Thickness (mm)} & \textbf{Parameters (mm)} \\
    \midrule
    Top Core Layer & $1 \times 1$ & $0.2$ & -- \\
    \midrule
    $\mu$Bump Layer & $1 \times 1$ & $0.04$ & $r=0.02$ \\
    \midrule
    Bottom Core Layer & $1 \times 1$ & $0.2$ & 
    \begin{tabular}{@{}c@{}}
      $r_{\mathrm{Cu}}=0.02$ \\
      $t_{ox}=0.01$
    \end{tabular} \\
    \midrule
    Base Die & $1 \times 1$ & $0.2$ & 
    \begin{tabular}{@{}c@{}}
      $r_{\mathrm{Cu}}=0.02$ \\
      $t_{ox}=0.01$
    \end{tabular} \\
    \midrule
    C4 Bump Layer & $1 \times 1$ & $0.1$ & $r=0.05$ \\
    \midrule
    Substrate & $10 \times 10$ & $1$ & -- \\
    \bottomrule
  \end{tabular}

  \vspace{4pt} 
  
  \begin{minipage}{\linewidth}
    \footnotesize
    \textit{Notes.}
    The TSVs form a $10 \times 10$ array aligned with the $\mu$bump distribution, and the C4 bumps form a $2 \times 2$ array.
  \end{minipage}
  \vspace*{-10pt}

\end{table}

For the industrial benchmarks, we construct multi-fidelity data by combining low-cost simulations from the EMT-based thermal-equivalent model with high-fidelity labels from detailed COMSOL FEM simulations~\cite{comsol}, where the TSV/\textmu{}bump geometries are explicitly resolved. 
For each industrial case, the training set contains 900 samples in total, with high- and low-fidelity samples mixed at a $1:3$ ratio, while the test set contains 100 high-fidelity COMSOL samples. 
This protocol evaluates Therm-FM on realistic thermal fields while limiting the amount of expensive high-fidelity supervision.

\subsection{Baselines}

We compare Therm-FM with representative learning-based thermal simulation methods, including FNO~\cite{FNO}, U-FNO~\cite{UFNO}, DeepOHeat~\cite{deepoheat}, ARO~\cite{wang2024aro}, SAU-FNO~\cite{saufno}, and T-Fusion~\cite{tfusion}. 
These baselines cover mainstream neural-operator, frequency-domain, and deep thermal-field prediction frameworks. 
Among them, SAU-FNO represents a recent SOTA baseline for learning-based thermal modeling.

This work focuses on operator learning from power distributions to temperature fields, where a trained model is expected to predict thermal responses for varying input samples. 
PINN-based methods~\cite{2024-pinn,thermal-pinn,2025-pinn-dac} are therefore not included as main quantitative baselines, since they primarily solve individual PDE instances or optimize for specific physical settings rather than learning a reusable input--output thermal operator.
For methods with publicly reported results under the same benchmark settings, we quote the official numbers. 
For missing settings, we reproduce the methods using available implementations or paper descriptions, following the official configurations whenever possible. 
All comparisons use the same evaluation metrics and corresponding benchmark resolutions. 
All baselines are trained using high-fidelity data only, following prior work. Therm-FM uses the same setting on HotSpot benchmarks, where no thermal-equivalent model is available, and adopts multi-fidelity training only on the industrial 3D-IC cases. As shown in Table~\ref{tab:hf_align}, multi-fidelity training achieves comparable accuracy to high-fidelity-only training while reducing high-fidelity data generation cost.

\subsection{Experimental Setup}

All experiments are conducted on an NVIDIA A100 GPU with 80GB memory and an Intel Xeon Gold 6246R CPU, using PyTorch 2.0.1 and CUDA 11.8. 
Unless otherwise specified, Therm-FM is trained with AdamW, a batch size of 40, and a weight decay of $1\times10^{-6}$. 
The pretrained backbone uses a learning rate of $5\times10^{-5}$, while the newly initialized embedding and recovery layers use $5\times10^{-4}$. 
During high-fidelity calibration, the learning rates are further reduced to stabilize fine-tuning under limited high-fidelity supervision. 
Hyperparameters are selected on the validation set, with Optuna used for automated search.

All input and output fields are standardized as
\(\hat{x} = \frac{x-\mu}{\sigma}\),
where $\mu$ and $\sigma$ are computed from the training set and fixed during validation and testing. 
The same normalization protocol is applied across benchmark settings and resolutions to ensure consistent evaluation.

Prediction accuracy is evaluated using \textbf{RMSE}, \textbf{MAPE}, \textbf{PAPE}, \textbf{Mean Absolute Error (Mean)}, and \textbf{Maximum Absolute Error (Max)}, where MAPE and PAPE are \textbf{reported in percentage (\%)} and PAPE denotes the peak absolute percentage error.
Computational efficiency is measured by \textbf{training time(Time)} and \textbf{peak GPU memory usage(GPU Mem)}. 
We further report \textbf{$\Delta$Mean} and \textbf{$\Delta$Max}, which denote the relative improvement ratios in Mean and Max errors over SAU-FNO under the same benchmark setting.
Larger $\Delta$ values indicate stronger relative error reduction.

\begin{table*}[htbp]
\caption{Comparison of Therm-FM with existing ML-based thermal modeling methods on HS-SC refine1 and refine2}
\vspace*{-0.05in}
\label{tab:baseline1}
\centering
\begin{tabular}{c|c|c|c|c|c|c|c|c|c|c}
\hline
\textbf{Method} & \textbf{Resolution} &
\textbf{RMSE} & \textbf{MAPE} & \textbf{PAPE} &
\textbf{Mean} & \textbf{$\Delta$Mean} &
\textbf{Max} & \textbf{$\Delta$Max} &
\textbf{Time(h)} & \textbf{GPU Mem} \\
\hline

FNO (ICLR'21)~\cite{FNO} & 55$\times$55 & 0.246 & 0.038 & 0.933 & 0.119 & 0.52 & 2.946 & 0.40 & \underline{0.16} & 0.53G \\
U-FNO (Adv. Water'23)~\cite{UFNO} & 55$\times$55 & 0.161 & 0.025 & 0.407 & 0.079 & 0.78 & 1.708 & 0.69 & 0.61 & 2.48G \\
DeepOHeat (DAC'23)~\cite{deepoheat} & 55$\times$55 & 0.263 & 0.041 & 0.975 & 0.124 & 0.50 & 3.430 & 0.35 & 0.83 & \underline{0.37G} \\
ARO (ICCAD'24)~\cite{wang2024aro} & 55$\times$55 & 0.212 & 0.034 & 0.830 & 0.120 & 0.52 & 2.768 & 0.43 & 0.65 & 2.36G \\
T-Fusion (ASPDAC'25)~\cite{tfusion} & 55$\times$55 & 0.547 & 0.146 & 1.746 & 0.279 & 0.22 & 5.430 & 0.22 & \textbf{0.03} & \textbf{0.19G} \\
\rowcolor{SotaColor}
SAU-FNO (DAC'25)~\cite{saufno} & 55$\times$55 & 0.119 & 0.018 & 0.334 & 0.062 & 1.00 & 1.185 & 1.00 & 2.11 & 7.74G \\
\rowcolor{OursColor}
Therm-FM-T (21M) & 55$\times$55 & 0.051 & 0.008 & 0.161 & 0.028 & 2.21 & 0.558 & 2.12 & 0.30 & 1.55G \\
\rowcolor{OursColor}
Therm-FM-B (158M) & 55$\times$55 & \underline{0.033} & \underline{0.005} & \underline{0.130} & \underline{0.017} & \underline{3.65} & \underline{0.435} & \underline{2.72} & 0.81 & 4.89G \\
\rowcolor{OursColor}
Therm-FM-L (629M) & 55$\times$55 & \textbf{0.027} & \textbf{0.004} & \textbf{0.116} & \textbf{0.014} & \textbf{4.43} & \textbf{0.384} & \textbf{3.09} & 1.48 & 15.14G \\
\hline

FNO (ICLR'21)~\cite{FNO} & 88$\times$88 & 0.183 & 0.028 & 0.591 & 0.096 & 0.51 & 2.190 & 0.40 & \underline{0.31} & 1.25G \\
U-FNO (Adv. Water'23)~\cite{UFNO} & 88$\times$88 & 0.130 & 0.019 & 0.292 & 0.075 & 0.65 & 1.117 & 0.78 & 1.32 & 5.28G \\
DeepOHeat (DAC'23)~\cite{deepoheat} & 88$\times$88 & 0.192 & 0.035 & 0.633 & 0.107 & 0.46 & 2.374 & 0.37 & 1.65 & \underline{0.89G} \\
ARO (ICCAD'24)~\cite{wang2024aro} & 88$\times$88 & 0.174 & 0.026 & 0.565 & 0.091 & 0.54 & 2.014 & 0.43 & 1.37 & 4.91G \\
T-Fusion (ASPDAC'25)~\cite{tfusion} & 88$\times$88 & 0.488 & 0.127 & 1.109 & 0.256 & 0.19 & 3.651 & 0.24 & \textbf{0.08} & \textbf{0.43G} \\
\rowcolor{SotaColor}
SAU-FNO (DAC'25)~\cite{saufno} & 88$\times$88 & 0.090 & 0.014 & 0.232 & 0.049 & 1.00 & 0.871 & 1.00 & 4.53 & 15.83G \\
\rowcolor{OursColor}
Therm-FM-T (21M) & 88$\times$88 & 0.040 & \underline{0.006} & 0.137 & 0.022 & 2.23 & 0.498 & 1.75 & 0.69 & 4.66G \\
\rowcolor{OursColor}
Therm-FM-B (158M) & 88$\times$88 & \underline{0.026} & \textbf{0.004} & \underline{0.083} & \underline{0.014} & \underline{3.50} & \underline{0.295} & \underline{2.95} & 1.52 & 12.45G \\
\rowcolor{OursColor}
Therm-FM-L (629M) & 88$\times$88 & \textbf{0.021} & \textbf{0.004} & \textbf{0.068} & \textbf{0.012} & \textbf{4.08} & \textbf{0.232} & \textbf{3.75} & 2.93 & 23.46G \\
\hline
\end{tabular}

\vspace{0.05in}
\footnotesize{
\textit{Notes.}
\colorbox{OursColor}{Therm-FM} and \colorbox{SotaColor}{SAU-FNO} highlighted.
Best results are \textbf{bolded}, second-best are \underline{underlined}.\\
$\Delta$Mean /$\Delta$Max denote improvement ratios versus SAU-FNO under same resolution.
}
\vspace*{-6pt}
\end{table*}

\begin{table*}[htbp]
\caption{Comparison of Therm-FM with existing ML-based thermal modeling methods on HS-OC refine1 and refine2}
\vspace*{-0.05in}
\label{tab:baseline_ind}
\centering

\begin{tabular}{c|c|c|c|c|c|c|c|c|c|c}
\hline
\textbf{Method} & \textbf{Resolution} &
\textbf{RMSE} & \textbf{MAPE} & \textbf{PAPE} &
\textbf{Mean} & \textbf{$\Delta$Mean} &
\textbf{Max} & \textbf{$\Delta$Max} &
\textbf{Time(h)} & \textbf{GPU Mem} \\
\hline

FNO (ICLR'21)~\cite{FNO} & 85$\times$85 & 0.697 & 0.205 & 0.910 & 0.502 & 0.31 & 2.781 & 0.52 & \underline{0.33} & 1.42G \\
U-FNO (Adv. Water'23)~\cite{UFNO} & 85$\times$85 & 0.356 & 0.089 & 0.623 & 0.249 & 0.63 & 1.480 & 0.98 & 1.47 & 5.63G \\
DeepOHeat (DAC'23)~\cite{deepoheat} & 85$\times$85 & 0.570 & 0.195 & 0.842 & 0.470 & 0.34 & 2.564 & 0.57 & 1.73 & \underline{0.87G} \\
ARO (ICCAD'24)~\cite{wang2024aro} & 85$\times$85 & 0.664 & 0.168 & 0.817 & 0.405 & 0.39 & 2.340 & 0.62 & 1.52 & 5.12G \\
T-Fusion (ASPDAC'25)~\cite{tfusion} & 85$\times$85 & 1.921 & 0.627 & 1.426 & 1.312 & 0.12 & 4.659 & 0.31 & \textbf{0.09} & \textbf{0.43G} \\
\rowcolor{SotaColor}
SAU-FNO (DAC'25)~\cite{saufno} & 85$\times$85 & 0.249 & 0.064 & 0.492 & 0.158 & 1.00 & 1.453 & 1.00 & 4.83 & 16.03G \\
\rowcolor{OursColor}
Therm-FM-T (21M) & 85$\times$85 & 0.117 & 0.023 & 0.334 & 0.077 & 2.05 & 1.198 & 1.21 & 0.70 & 4.67G \\
\rowcolor{OursColor}
Therm-FM-B (158M) & 85$\times$85 & \underline{0.069} & \underline{0.013} & \underline{0.198} & \underline{0.045} & \underline{3.51} & \underline{0.708} & \underline{2.05} & 1.80 & 12.45G \\
\rowcolor{OursColor}
Therm-FM-L (629M) & 85$\times$85 & \textbf{0.035} & \textbf{0.007} & \textbf{0.097} & \textbf{0.023} & \textbf{6.87} & \textbf{0.346} & \textbf{4.20} & 4.40 & 23.42G \\

FNO (ICLR'21)~\cite{FNO} & 151$\times$151 & 0.707 & 0.212 & 1.025 & 0.505 & 0.34 & 3.029 & 0.47 & \underline{0.42} & 2.12G \\
U-FNO (Adv. Water'23)~\cite{UFNO} & 151$\times$151 & 0.440 & 0.125 & 0.668 & 0.277 & 0.62 & 1.841 & 0.78 & 1.82 & 8.23G \\
DeepOHeat (DAC'23)~\cite{deepoheat} & 151$\times$151 & 0.712 & 0.237 & 0.981 & 0.539 & 0.32 & 3.183 & 0.45 & 2.13 & \underline{1.22G} \\
ARO (ICCAD'24)~\cite{wang2024aro} & 151$\times$151 & 0.681 & 0.185 & 0.951 & 0.487 & 0.35 & 2.893 & 0.50 & 1.93 & 7.53G \\
T-Fusion (ASPDAC'25)~\cite{tfusion} & 151$\times$151 & 2.182 & 0.736 & 1.623 & 1.462 & 0.12 & 5.357 & 0.27 & \textbf{0.11} & \textbf{0.63G} \\
\rowcolor{SotaColor}
SAU-FNO (DAC'25)~\cite{saufno} & 151$\times$151 & 0.296 & 0.069 & 0.481 & 0.172 & 1.00 & 1.435 & 1.00 & 6.23 & 28.53G \\
\rowcolor{OursColor}
Therm-FM-T (21M) & 151$\times$151 & 0.093 & \underline{0.018} & 0.345 & \underline{0.060} & \underline{2.87} & 1.240 & 1.16 & 1.63 & 15.72G \\
\rowcolor{OursColor}
Therm-FM-B (158M) & 151$\times$151 & \underline{0.088} & \underline{0.018} & \underline{0.241} & 0.061 & 2.82 & \underline{0.863} & \underline{1.66} & 3.56 & 38.13G \\
\rowcolor{OursColor}
Therm-FM-L (629M) & 151$\times$151 & \textbf{0.069} & \textbf{0.015} & \textbf{0.196} & \textbf{0.049} & \textbf{3.51} & \textbf{0.701} & \textbf{2.05} & 6.75 & 60.77G \\
\hline
\end{tabular}

\vspace{0.05in}
\vspace{-8pt}
\end{table*}

\begin{figure*}[htbp]
    \vspace*{-4pt}
    \centering
    \includegraphics[width=\textwidth]{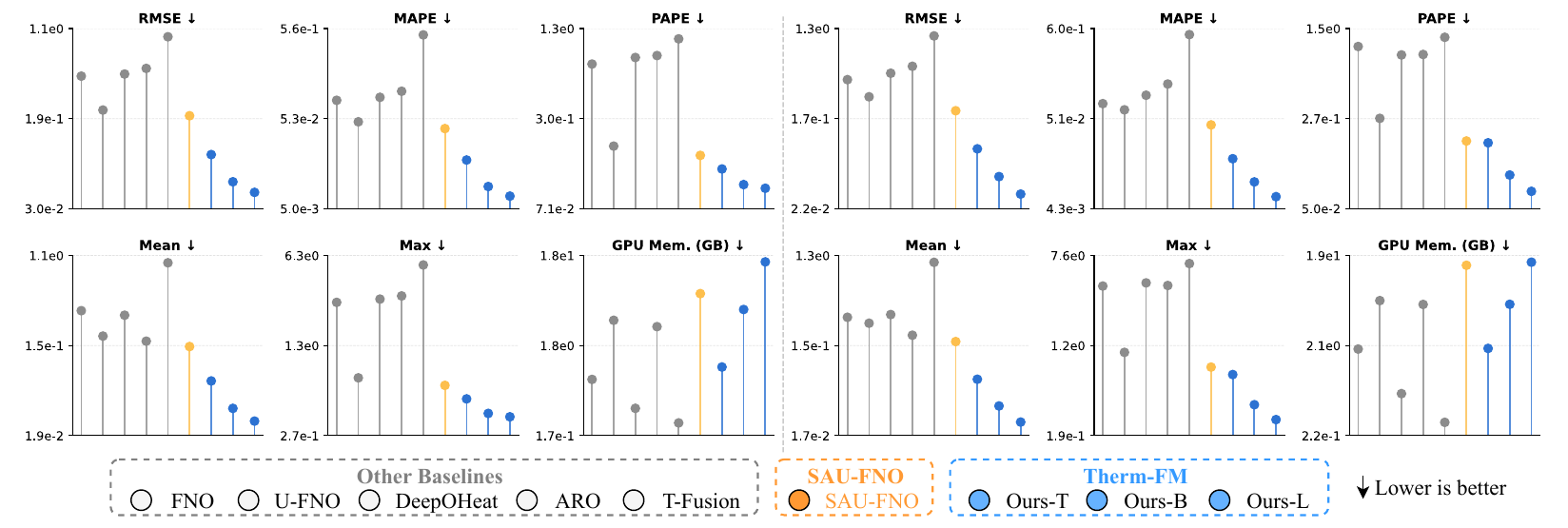}
  \caption{
Comparison with existing methods on the HS-QC case at $40\times40$ and $64\times64$ resolutions.
Each subpanel reports one metric, including RMSE, MAPE, PAPE, Mean error, Max error, and GPU memory.
Gray markers denote prior baselines in the legend order, orange denotes SAU-FNO, and blue denotes Therm-FM variants.
Lower values are better for all metrics.
}
    \label{Fig:baseline1}
    \vspace*{-10pt} 
\end{figure*}

\subsection{Thermal-Equivalent Model Validation}

We validate the proposed thermal-equivalent model at both the layer and system levels to assess whether it can provide reliable low-fidelity data for multi-fidelity training.

\subsubsection{Layer-Level Accuracy and Geometric Robustness}
We first evaluate a local TSV test structure associated with the IND-8C package stack under a uniform step power input of $0.2\,\mathrm{W}$. 
As shown in Fig.~5(a), the transient temperature predicted by the EMT-based thermal-equivalent model closely matches the high-fidelity COMSOL result over $0$--$1.5\,\mathrm{s}$, indicating that the equivalent heat capacity preserves the thermal time response. 
We further vary the TSV radius and pitch, increasing the TSV volume fraction from $2.5\%$ to $50\%$. 
Fig.~5(b) shows that the maximum steady-state relative error remains below $1\%$ within the practical TSV density range below $20\%$, and is about $3.5\%$ even in the high-density $50\%$ case.

\subsubsection{System-Level Accuracy and Efficiency}
On the complete IND-8C benchmark, explicitly resolving TSV and \textmu{}bump structures in COMSOL requires over $28.5$M mesh elements, whereas the thermal-equivalent model uses only $8{,}405$ elements, reducing mesh complexity by approximately $3400\times$. 
For steady-state simulation, the solution time decreases from $1766\,\mathrm{s}$ to $6\,\mathrm{s}$, achieving a $294\times$ speedup with a temperature-field relative error of $0.63\%$. 
For transient simulation, the solution time decreases from $4261\,\mathrm{s}$ to $8\,\mathrm{s}$, achieving a $532\times$ speedup with a maximum relative temperature error of $1.12\%$. 
These results support the use of the thermal-equivalent model as a reliable low-cost source of approximate data for Therm-FM training.

\begin{figure}[tbp]
    \centering
    \begin{minipage}{0.48\linewidth}
        \centering
        \includegraphics[width=\linewidth]{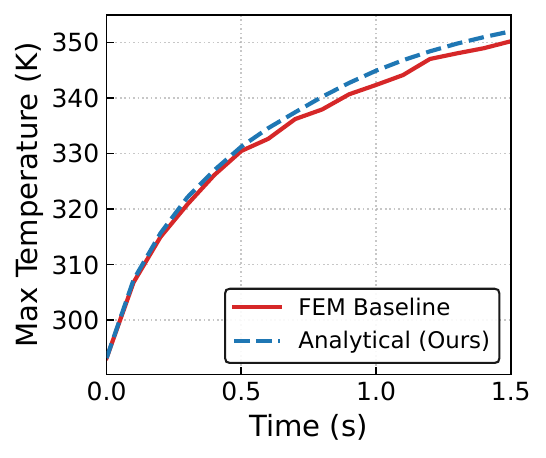}
        \centerline{\footnotesize (a) Transient Response}
        \label{fig:validation_transient}
    \end{minipage}
    \hfill
    \begin{minipage}{0.48\linewidth}
        \centering
        \includegraphics[width=\linewidth]{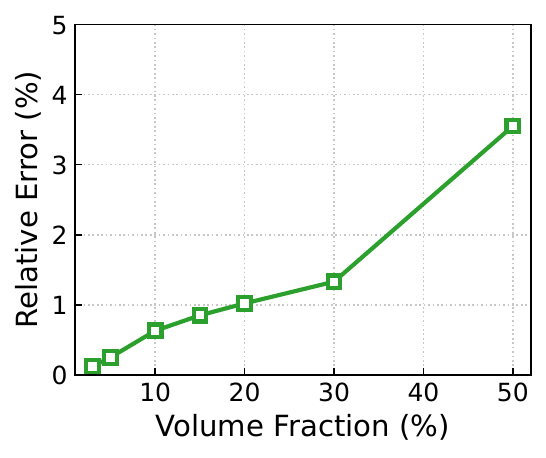}
        \centerline{\footnotesize (b) Parameter Sweep}
        \label{fig:validation_sweep}
    \end{minipage}
    \caption{Validation of the thermal-equivalent model on a TSV layer. 
    (a) Transient temperature response under a step power input. 
    (b) Maximum steady-state relative error under different TSV volume fractions.}
    \label{fig:local_validation}
    
\end{figure}

\subsection{Steady-State Thermal Prediction}

We first evaluate Therm-FM on steady-state thermal prediction across both public HotSpot benchmarks and industrial 3D-IC package cases.

\textbf{HotSpot-based benchmarks.}
As shown in Fig.~\ref{Fig:baseline1} and Tables~\ref{tab:baseline1}--\ref{tab:baseline_ind}, Therm-FM consistently achieves lower prediction error than existing learning-based thermal simulation methods across different HotSpot configurations, spatial resolutions, and model scales.
On HS-SC, Therm-FM-L reduces the RMSE of SAU-FNO from $0.119$ to $0.027$ at $55\times55$ resolution and from $0.090$ to $0.021$ at $88\times88$ resolution, corresponding to about $4.4\times$ and $4.3\times$ lower RMSE, respectively.
On HS-OC, which has the largest core count and strongest inter-die thermal coupling among the HotSpot cases, Therm-FM-L reduces the RMSE of SAU-FNO from $0.249$ to $0.035$ at $85\times85$ resolution and from $0.296$ to $0.069$ at $151\times151$ resolution, corresponding to about $7.1\times$ and $4.3\times$ error reductions, respectively.
Notably, at $85\times85$ resolution, Therm-FM-L also reduces the PAPE from $0.492$ to $0.097$ and the Max error from $1.453$ to $0.346$, corresponding to about $5.1\times$ and $4.2\times$ improvements, respectively.
These results demonstrate stronger robustness on hotspot-sensitive regions in addition to improved global temperature-field reconstruction.

Across most settings, prediction error consistently decreases as model scale increases from Therm-FM-T to Therm-FM-L, indicating that Therm-FM makes better use of increased model capacity than other methods.
One exception is observed on HS-OC at $151\times151$ resolution, where Therm-FM-B achieves a Mean error of $0.061$, marginally higher than Therm-FM-T ($0.060$).
We attribute this to insufficient training convergence of the Base model at this resolution: the $151\times151$ grid produces a substantially larger output space, and the Base configuration may require more training epochs or a larger batch size to fully exploit its capacity advantage.
This isolated non-monotonic point does not affect the overall scaling trend observed across the remaining benchmarks.
The consistent improvements in PAPE and Max error further show that Therm-FM improves not only global temperature-field reconstruction but also hotspot-sensitive prediction.

\begin{table*}[!htbp]
\caption{Comparison of Therm-FM with existing ML-based thermal modeling methods on IND-8C(upper) and IND-32C(below)}
\vspace*{-0.05in}
\label{tab:baseline2}
\centering

\begin{tabular}{c|c|c|c|c|c|c|c|c|c|c}
\hline
\textbf{Method} & \textbf{Resolution} &
\textbf{RMSE} & \textbf{MAPE} & \textbf{PAPE} &
\textbf{Mean} & \textbf{$\Delta$Mean} &
\textbf{Max} & \textbf{$\Delta$Max} &
\textbf{Time(h)} & \textbf{GPU Mem} \\
\hline
FNO (ICLR'21)~\cite{FNO} & 101$\times$101 & 0.157 & 0.035 & 0.374 & 0.124 & 0.64 & 1.358 & 0.60 & 0.28 & 1.96G \\
U-FNO (Adv. Water'23)~\cite{UFNO} & 101$\times$101 & 0.116 & 0.019 & 0.233 & 0.094 & 0.84 & 0.866 & 0.94 & 1.20 & 7.97G \\
DeepOHeat (DAC'23)~\cite{deepoheat} & 101$\times$101 & 0.173 & 0.047 & 0.461 & 0.152 & 0.52 & 1.494 & 0.54 & 1.46 & \underline{1.20G} \\
ARO (ICCAD'24)~\cite{wang2024aro} & 101$\times$101 & 0.135 & 0.031 & 0.308 & 0.098 & 0.81 & 1.157 & 0.70 & 1.24 & 7.13G \\
T-Fusion (ASPDAC'25)~\cite{tfusion} & 101$\times$101 & 0.702 & 0.136 & 0.765 & 0.533 & 0.15 & 3.409 & 0.24 & \textbf{0.07} & \textbf{0.63G} \\
\rowcolor{SotaColor}
SAU-FNO (DAC'25)~\cite{saufno} & 101$\times$101 & 0.104 & 0.015 & 0.192 & 0.079 & 1.00 & 0.813 & 1.00 & 3.95 & 21.80G \\
\rowcolor{OursColor}
Therm-FM-T (21M) & 101$\times$101 & 0.032 & 0.007 & \underline{0.111} & 0.024 & 3.29 & \underline{0.393} & \underline{2.07} & \underline{0.23} & 5.24G \\
\rowcolor{OursColor}
Therm-FM-B (158M) & 101$\times$101 & \underline{0.019} & \underline{0.004} & 0.133 & \underline{0.013} & \underline{6.08} & 0.474 & 1.72 & 0.62 & 14.63G \\
\rowcolor{OursColor}
Therm-FM-L (629M) & 101$\times$101 & \textbf{0.011} & \textbf{0.002} & \textbf{0.046} & \textbf{0.008} & \textbf{9.88} & \textbf{0.162} & \textbf{5.02} & 1.45 & 22.60G \\
\hline

FNO (ICLR'21)~\cite{FNO} & 101$\times$101 & 0.165 & 0.032 & 0.317 & 0.126 & 0.67 & 1.096 & 0.64 & 0.28 & 1.96G \\
U-FNO (Adv. Water'23)~\cite{UFNO} & 101$\times$101 & 0.124 & 0.025 & 0.212 & 0.091 & 0.93 & 0.768 & 0.92 & 1.20 & 7.91G \\
DeepOHeat (DAC'23)~\cite{deepoheat} & 101$\times$101 & 0.187 & 0.048 & 0.428 & 0.155 & 0.55 & 1.187 & 0.59 & 1.45 & \underline{1.19G} \\
ARO (ICCAD'24)~\cite{wang2024aro} & 101$\times$101 & 0.139 & 0.027 & 0.305 & 0.103 & 0.83 & 0.915 & 0.77 & 1.25 & 7.06G \\
T-Fusion (ASPDAC'25)~\cite{tfusion} & 101$\times$101 & 0.838 & 0.144 & 0.850 & 0.655 & 0.13 & 3.772 & 0.19 & \textbf{0.07} & \textbf{0.64G} \\
\rowcolor{SotaColor}
SAU-FNO (DAC'25)~\cite{saufno} & 101$\times$101 & 0.096 & 0.019 & 0.186 & 0.085 & 1.00 & 0.706 & 1.00 & 3.90 & 22.10G \\
\rowcolor{OursColor}
Therm-FM-T (21M) & 101$\times$101 & 0.030 & 0.006 & \underline{0.083} & 0.022 & 3.86 & \underline{0.293} & \underline{2.41} & \underline{0.23} & 5.24G \\
\rowcolor{OursColor}
Therm-FM-B (158M) & 101$\times$101 & \underline{0.018} & \underline{0.004} & 0.094 & \underline{0.013} & \underline{6.54} & 0.336 & 2.10 & 0.57 & 14.63G \\
\rowcolor{OursColor}
Therm-FM-L (629M) & 101$\times$101 & \textbf{0.010} & \textbf{0.002} & \textbf{0.037} & \textbf{0.008} & \textbf{10.62} & \textbf{0.131} & \textbf{5.39} & 1.42 & 22.60G \\
\hline
\end{tabular}

\vspace{0.05in}
\vspace{-8pt}
\end{table*}

\textbf{Industrial benchmarks.}
We further evaluate Therm-FM on IND-8C and IND-32C, which contain multi-layer stacked packaging structures with TSV/\textmu{}bump arrays and industrial-package thermal interactions.
As reported in Table~\ref{tab:baseline2}, Therm-FM-L reduces RMSE by about $9.5\times$--$9.6\times$ over SAU-FNO on the two industrial cases and achieves up to a $10.62\times$ reduction in Mean error, outperforming all other baselines including FNO, U-FNO, DeepOHeat, ARO, and T-Fusion.
Despite the denser interconnect structures and smaller training set, Therm-FM achieves Mean and Max errors comparable to, or lower than, those on the HotSpot cases, demonstrating its ability to handle complex package-level thermal prediction.

On IND-8C, Therm-FM-L reduces the RMSE of SAU-FNO from $0.104$ to $0.011$, corresponding to a $9.45\times$ error reduction, decreases the Mean error from $0.079$ to $0.008$ with a $9.88\times$ reduction, and reduces the Max error from $0.813$ to $0.162$ with a $5.02\times$ reduction. 
Even Therm-FM-T already outperforms all previous methods with an RMSE of $0.032$ while using only $5.24$GB GPU memory.

On IND-32C, Therm-FM-L reduces the RMSE of SAU-FNO from $0.096$ to $0.010$, corresponding to a $9.60\times$ error reduction, lowers the PAPE from $0.186$ to $0.037$, decreases the Mean error from $0.085$ to $0.008$ with a $10.62\times$ reduction, and reduces the Max error from $0.706$ to $0.131$ with a $5.39\times$ reduction. 

The qualitative results in Fig.~\ref{Fig:steady_visualization} show that Therm-FM accurately preserves both global temperature distributions and local hotspot structures across different package layouts, with only small errors near sharp thermal boundaries.

\subsection{Transient Thermal Prediction}

\begin{table*}[htbp]
\caption{Transient thermal prediction results on HS-SC, HS-QC, and HS-OC benchmarks.}
\vspace*{-0.05in}
\label{tab:baseline_multi}
\centering

\begin{tabular}{c|c|c|c|c|c|c|c|c|c|c}
\hline
\textbf{Method} & \textbf{Resolution} &
\textbf{RMSE} & \textbf{MAPE} & \textbf{PAPE} &
\textbf{Mean} & \textbf{$\Delta$Mean} &
\textbf{Max} & \textbf{$\Delta$Max} &
\textbf{Time(h)} & \textbf{GPU Mem} \\
\hline

FNO (ICLR'21)~\cite{FNO} & 88$\times$88$\times$9 & 0.163 & 0.034 & 0.308 & 0.091 & 0.42 & 0.820 & 0.66 & \underline{0.46} & 2.12G \\
U-FNO (Adv. Water'23)~\cite{UFNO} & 88$\times$88$\times$9 & 0.108 & 0.015 & 0.210 & 0.056 & 0.68 & 0.595 & 0.91 & 1.63 & 6.23G \\
DeepOHeat (DAC'23)~\cite{deepoheat} & 88$\times$88$\times$9 & 0.151 & 0.027 & 0.293 & 0.082 & 0.46 & 0.777 & 0.70 & 1.93 & \underline{1.02G} \\
ARO (ICCAD'24)~\cite{wang2024aro} & 88$\times$88$\times$9 & 0.135 & 0.029 & 0.289 & 0.067 & 0.57 & 0.753 & 0.72 & 1.72 & 5.63G \\
T-Fusion (ASPDAC'25)~\cite{tfusion} & 88$\times$88$\times$9 & 0.660 & 0.122 & 0.727 & 0.336 & 0.11 & 2.285 & 0.24 & \textbf{0.11} & \textbf{0.53G} \\
\rowcolor{SotaColor}
SAU-FNO (DAC'25)~\cite{saufno} & 88$\times$88$\times$9 & 0.077 & 0.012 & 0.193 & 0.038 & 1.00 & 0.543 & 1.00 & 4.93 & 16.83G \\
\rowcolor{OursColor}
Therm-FM-T (21M) & 88$\times$88$\times$9 & 0.018 & \underline{0.002} & 0.091 & 0.008 & 4.75 & 0.305 & 1.78 & 0.92 & 6.33G \\
\rowcolor{OursColor}
Therm-FM-B (158M) & 88$\times$88$\times$9 & \underline{0.011} & \underline{0.002} & \underline{0.055} & \underline{0.005} & \underline{7.60} & \underline{0.185} & \underline{2.94} & 1.64 & 14.09G \\
\rowcolor{OursColor}
Therm-FM-L (629M) & 88$\times$88$\times$9 & \textbf{0.009} & \textbf{0.001} & \textbf{0.042} & \textbf{0.004} & \textbf{9.50} & \textbf{0.140} & \textbf{3.88} & 3.42 & 25.21G \\
\hline

FNO (ICLR'21)~\cite{FNO} & 64$\times$64$\times$5 & 0.295 & 0.067 & 0.618 & 0.168 & 0.48 & 1.452 & 0.59 & \underline{0.21} & 0.83G \\
U-FNO (Adv. Water'23)~\cite{UFNO} & 64$\times$64$\times$5 & 0.173 & 0.038 & 0.426 & 0.104 & 0.77 & 1.074 & 0.80 & 0.87 & 2.63G \\
DeepOHeat (DAC'23)~\cite{deepoheat} & 64$\times$64$\times$5 & 0.276 & 0.059 & 0.604 & 0.146 & 0.55 & 1.617 & 0.53 & 1.03 & \underline{0.47G} \\
ARO (ICCAD'24)~\cite{wang2024aro} & 64$\times$64$\times$5 & 0.241 & 0.034 & 0.552 & 0.129 & 0.62 & 1.503 & 0.57 & 0.94 & 2.33G \\
T-Fusion (ASPDAC'25)~\cite{tfusion} & 64$\times$64$\times$5 & 0.812 & 0.203 & 1.271 & 0.496 & 0.16 & 3.412 & 0.25 & \textbf{0.06} & \textbf{0.23G} \\
\rowcolor{SotaColor}
SAU-FNO (DAC'25)~\cite{saufno} & 64$\times$64$\times$5 & 0.127 & 0.028 & 0.347 & 0.080 & 1.00 & 0.861 & 1.00 & 2.43 & 6.83G \\
\rowcolor{OursColor}
Therm-FM-T (21M) & 64$\times$64$\times$5 & 0.040 & 0.008 & 0.115 & 0.025 & 3.20 & 0.377 & 2.28 & 0.35 & 1.98G \\
\rowcolor{OursColor}
Therm-FM-B (158M) & 64$\times$64$\times$5 & \underline{0.025} & \underline{0.005} & \underline{0.069} & \underline{0.016} & \underline{5.00} & \underline{0.225} & \underline{3.83} & 0.83 & 5.74G \\
\rowcolor{OursColor}
Therm-FM-L (629M) & 64$\times$64$\times$5 & \textbf{0.016} & \textbf{0.003} & \textbf{0.051} & \textbf{0.011} & \textbf{7.27} & \textbf{0.168} & \textbf{5.12} & 1.67 & 15.01G \\
\hline

FNO (ICLR'21)~\cite{FNO} & 151$\times$151$\times$9 & 0.834 & 0.162 & 1.861 & 0.503 & 0.43 & 4.318 & 0.66 & \underline{0.62} & 3.03G \\
U-FNO (Adv. Water'23)~\cite{UFNO} & 151$\times$151$\times$9 & 0.578 & 0.104 & 1.274 & 0.330 & 0.65 & 3.623 & 0.79 & 2.53 & 10.23G \\
DeepOHeat (DAC'23)~\cite{deepoheat} & 151$\times$151$\times$9 & 0.816 & 0.143 & 1.947 & 0.451 & 0.47 & 4.946 & 0.58 & 2.83 & \underline{1.63G} \\
ARO (ICCAD'24)~\cite{wang2024aro} & 151$\times$151$\times$9 & 0.684 & 0.128 & 1.618 & 0.397 & 0.54 & 4.287 & 0.67 & 2.63 & 9.43G \\
T-Fusion (ASPDAC'25)~\cite{tfusion} & 151$\times$151$\times$9 & 2.183 & 0.439 & 3.542 & 1.221 & 0.18 & 9.684 & 0.30 & \textbf{0.13} & \textbf{0.83G} \\
\rowcolor{SotaColor}
SAU-FNO (DAC'25)~\cite{saufno} & 151$\times$151$\times$9 & 0.409 & 0.084 & 1.062 & 0.214 & 1.00 & 2.869 & 1.00 & 7.03 & 30.03G \\
\rowcolor{OursColor}
Therm-FM-T (21M) & 151$\times$151$\times$9 & 0.125 & 0.019 & 0.842 & 0.061 & 3.51 & 3.138 & 0.91 & 1.70 & 17.84G \\
\rowcolor{OursColor}
Therm-FM-B (158M) & 151$\times$151$\times$9 & \underline{0.094} & \underline{0.015} & \underline{0.558} & \underline{0.049} & \underline{4.37} & \underline{2.075} & \underline{1.38} & 3.30 & 39.53G \\
\rowcolor{OursColor}
Therm-FM-L (629M) & 151$\times$151$\times$9 & \textbf{0.060} & \textbf{0.009} & \textbf{0.396} & \textbf{0.030} & \textbf{7.13} & \textbf{1.467} & \textbf{1.96} & 5.78 & 62.20G \\
\hline
\end{tabular}

\vspace{0.05in}

\vspace{-8pt}
\end{table*}

We further evaluate Therm-FM on transient thermal prediction across HS-SC, HS-QC, and HS-OC.
Compared with steady-state prediction, transient modeling requires capturing both spatial heat diffusion and temporal temperature evolution.
As reported in Table~\ref{tab:baseline_multi}, Therm-FM-L achieves the best prediction accuracy across all three transient benchmarks.

On HS-SC ($88\times88\times9$), Therm-FM-L reduces the RMSE of SAU-FNO from $0.077$ to $0.009$, corresponding to an $8.6\times$ error reduction.
It also decreases the PAPE from $0.193$ to $0.042$ and the Max error from $0.543$ to $0.140$, showing improved accuracy in transient hotspot prediction.
On HS-QC ($64\times64\times5$), Therm-FM-L reduces the RMSE from $0.127$ to $0.016$ and the Max error from $0.861$ to $0.168$, corresponding to $7.9\times$ and $5.1\times$ reductions, respectively.
On HS-OC ($151\times151\times9$), which has the largest spatial scale and strongest spatiotemporal coupling among the transient benchmarks, Therm-FM-L reduces the RMSE from $0.409$ to $0.060$ and the Max error from $2.869$ to $1.467$, corresponding to $6.8\times$ and $1.96\times$ reductions, respectively. 

One exception is observed for Therm-FM-T on HS-OC at $151\times151\times9$ resolution, where the Max error ($3.138$) is higher than that of SAU-FNO ($2.869$).
This is mainly due to the limited capacity of the Tiny configuration (21M parameters) under large-scale spatiotemporal inputs: the $151\times151\times9$ output grid contains nearly three times more tokens than the $88\times88\times9$ setting, making it harder for the small backbone to resolve sharp thermal gradients near hotspot boundaries.
The Base and Large configurations do not exhibit this issue and both reduce the Max error below SAU-FNO, indicating that the anomaly is capacity-limited rather than a systematic failure of the framework.

Fig.~\ref{Fig:transient_visualization} visualizes representative transient predictions on HS-SC and HS-OC.
For space efficiency, five representative frames are shown from each nine-frame sequence.
Therm-FM closely follows the ground-truth temperature evolution, including hotspot formation, heat diffusion, and cross-layer thermal propagation.

\subsection{Data Efficiency Analysis}

\begin{figure*}[htbp]
    \centering
    \includegraphics[width=\textwidth]{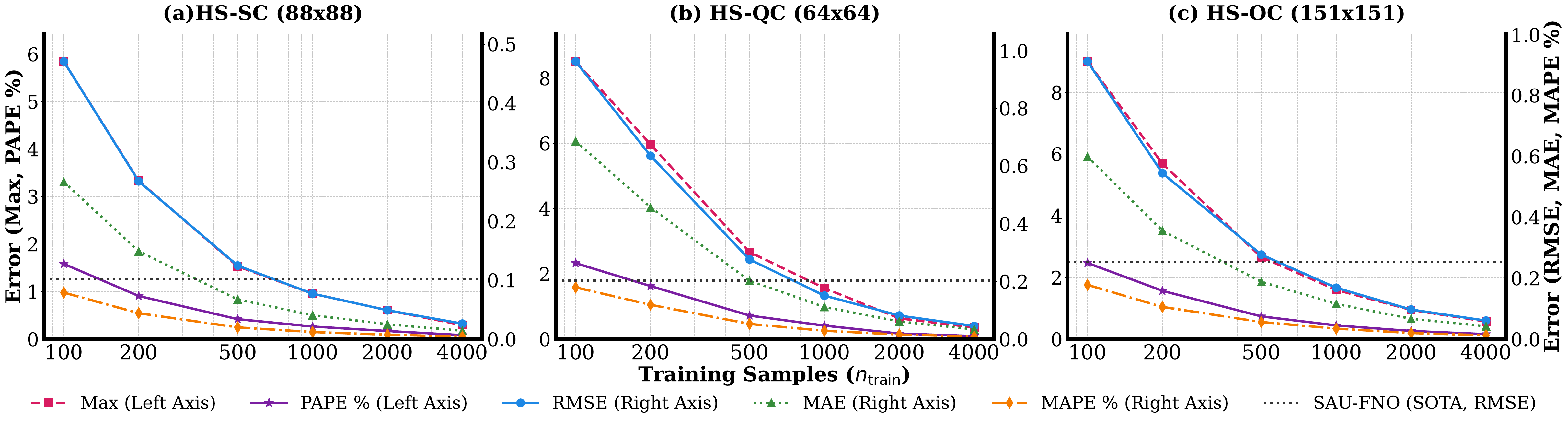}
    \vspace*{-20pt}
    \caption{
        Training-sample sensitivity of Therm-FM on HS-SC, HS-QC, and HS-OC. 
        The gray dashed line denotes the full-data SAU-FNO RMSE baseline. 
        All results are obtained with the \textbf{Therm-FM-B} configuration.
    }
    \label{fig:sensitivity_analysis_combined}
    \vspace*{-6pt}
\end{figure*}

To evaluate data efficiency, we vary the number of training samples on HS-SC, HS-QC, and HS-OC, covering spatial resolutions from $64\times64$ to $151\times151$.
As shown in Fig.~\ref{fig:sensitivity_analysis_combined}, Therm-FM consistently reduces RMSE, Mean(MAE), MAPE, PAPE, and Max error as more training samples are used.
The improvement is most pronounced in the low-data regime of 100--1000 samples, indicating that the pretrained PDE foundation model provides useful physical priors for rapid adaptation.

Therm-FM approaches or surpasses the full-data SAU-FNO baseline with only about 500--1000 training samples across the three designs.
After this range, the error reduction becomes more gradual, suggesting that the dominant thermal patterns have already been captured.
These results demonstrate that Therm-FM achieves strong accuracy with only a fraction of the full training set, which is critical for 3D-IC thermal modeling where high-fidelity data generation is expensive.

\subsection{Cross-Chip Generalization}

To evaluate cross-chip generalization, we conduct bidirectional transfer experiments between two industrial 3D-IC package cases, IND-8C and IND-32C. 
Specifically, models are first trained on the full training set of one dataset and then fine-tuned with only a few samples ($10$, $30$, $50$, $80$, and $100$) from the other dataset. 
This setting reflects practical design scenarios where historical simulation data are available for existing chip architectures, while only limited high-fidelity data can be generated for a new design.

\begin{figure*}[!htbp]
    \centering
    \includegraphics[width=0.98\textwidth]{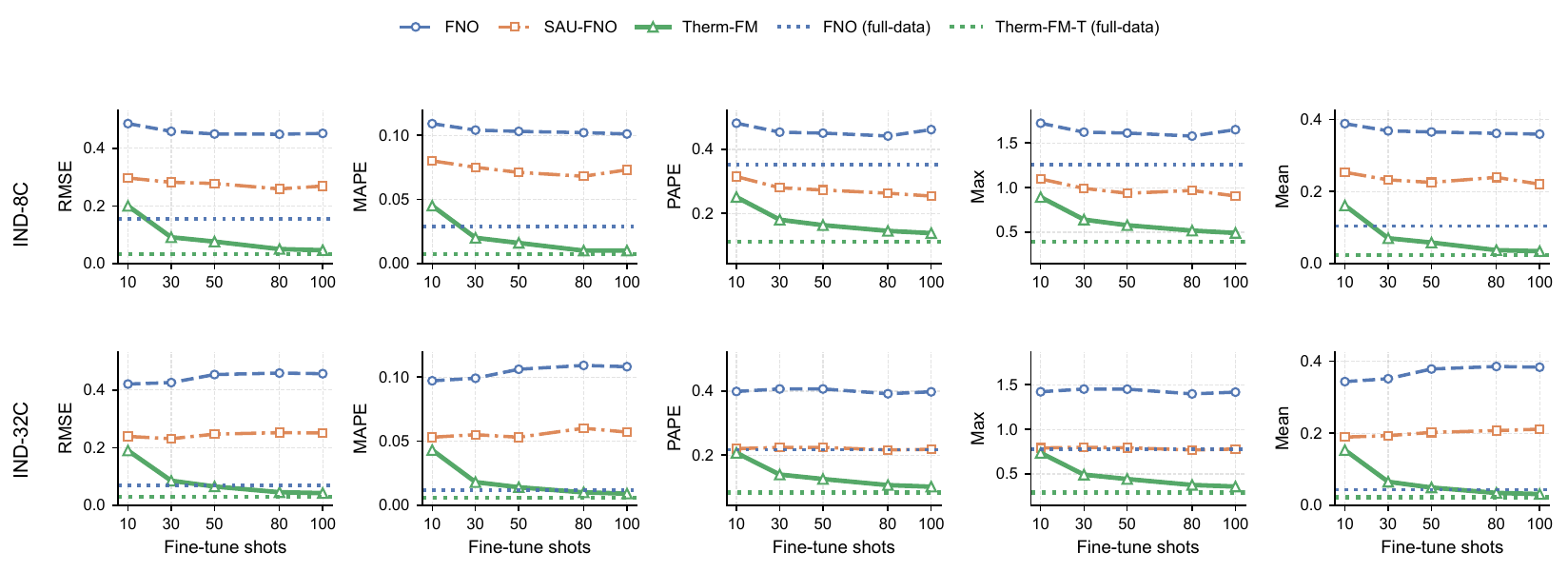}
    \caption{
    Few-shot cross-chip adaptation on the IND-8C and IND-32C cases.
    Models are trained on one industrial package case and fine-tuned with limited target samples from the other case.
    Each column reports one lower-is-better metric, including RMSE, MAPE, PAPE, Max, and Mean.
    Dashed horizontal lines denote full-data reference results obtained using all target training samples, while solid curves show few-shot adaptation performance.
    }
    \label{fig:fewshot_fulltrain_scaling}
\end{figure*}

As shown in Fig.~\ref{fig:fewshot_fulltrain_scaling}, FNO and SAU-FNO show limited transferability under limited-data adaptation in both directions. 
Even with 100 fine-tuning samples, their errors remain noticeably higher than the corresponding full-training baselines using all 900 samples, and increasing the number of fine-tuning samples brings only marginal improvements. 
This suggests that methods trained from scratch have limited cross-chip transferability and still require sufficient target-domain data to reach satisfactory accuracy. In contrast, Therm-FM demonstrates substantially stronger cross-chip adaptation. 
With only 10--30 fine-tuning samples, Therm-FM already achieves large error reductions and surpasses the full-data SAU-FNO baseline. 
With 50--80 samples, it further approaches the performance of full-data Therm-FM training using 900 samples. 
The consistent trends across both transfer directions indicate that Therm-FM captures reusable thermal-diffusion priors across chip architectures.
These results validate the effectiveness of foundation-model-based adaptation for 3D-IC thermal prediction with limited fine-tuning data.

\begin{figure*}[!htbp]
    \centering
    \includegraphics[width=0.98\textwidth]{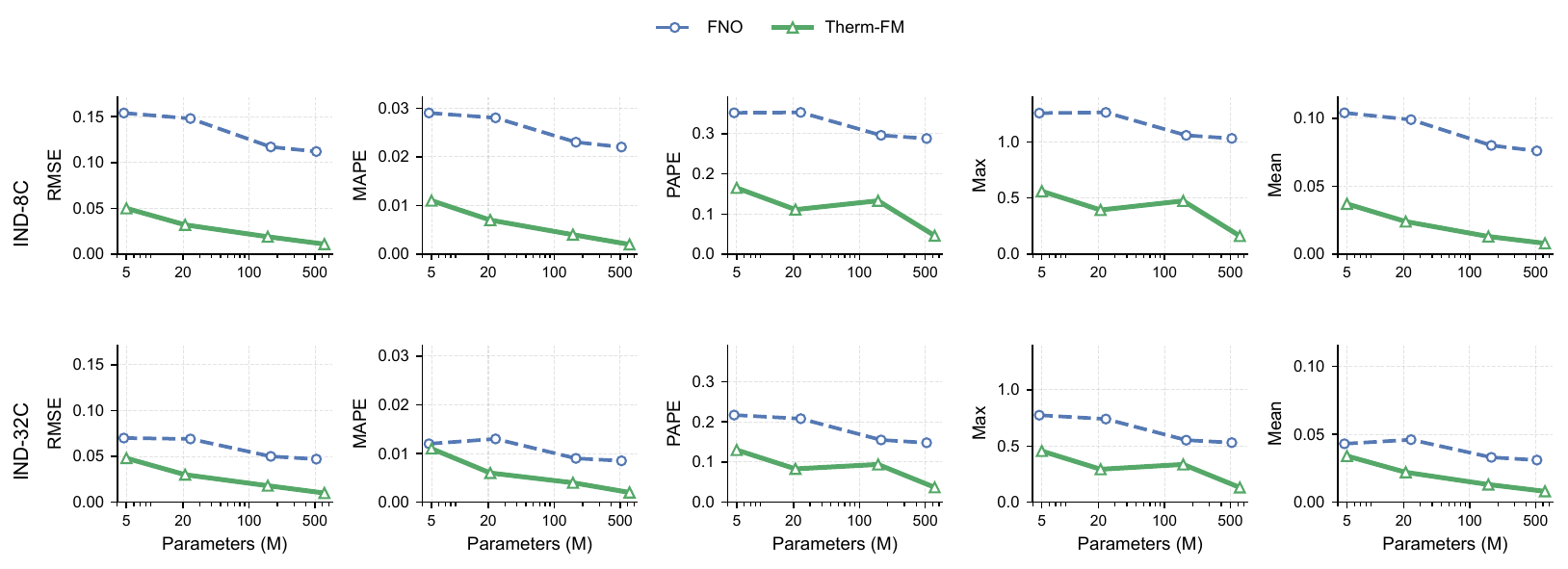}
    \caption{
    Performance trends with respect to model parameters on IND-8C and IND-32C cases.
    Each column reports one metric, and the two rows correspond to different settings.
    Our method consistently achieves lower errors than FNO across parameter scales.
    }
    \label{fig:param_metric_scaling}
\end{figure*}

\subsection{Ablation Study}

We further conduct ablation studies on pretrained initialization, model scaling, and multi-fidelity training in Therm-FM, showing their effects on data efficiency, accuracy, and data-generation cost.

\textbf{Effect of pretrained initialization.}
We first evaluate the impact of pretrained weights by comparing models trained from random initialization with those fine-tuned from the PDE foundation model.
As shown in Table~\ref{tab:ablation_pretrain}, on HS-QC ($64\times64$), all scratch variants already achieve lower RMSE than SAU-FNO, suggesting that the Therm-FM architecture itself provides a strong capacity advantage on relatively simpler benchmarks.

\begin{table}[htbp]
\caption{Ablation on pretrained initialization and model scaling.}
\vspace*{-0.05in}
\label{tab:ablation_pretrain}
\centering
\footnotesize
\setlength{\tabcolsep}{4.5pt}
\renewcommand{\arraystretch}{1.05}

\begin{tabular}{c|c|c|c|c|c|c}
\hline
\textbf{Model} & \textbf{Init.} &
\textbf{RMSE} & \textbf{MAPE} & \textbf{PAPE} &
\textbf{Max} & \textbf{Mean} \\
\hline

\multicolumn{7}{c}{\textit{HS-QC Refine2 ($64\times64$)}} \\
\hline
SAU-FNO & -- & 0.203 & 0.043 & 0.179 & 0.773 & 0.162 \\

\rowcolor{SotaColor}
Therm-FM-T & Scratch & 0.113 & 0.022 & 0.260 & 1.000 & 0.080 \\
\rowcolor{SotaColor}
Therm-FM-B & Scratch & 0.065 & 0.013 & 0.138 & 0.527 & 0.048 \\
\rowcolor{SotaColor}
Therm-FM-L & Scratch & 0.048 & 0.009 & 0.090 & 0.350 & 0.036 \\

\rowcolor{OursColor}
Therm-FM-T & Pretrained & 0.086 & 0.017 & 0.173 & 0.663 & 0.065 \\
\rowcolor{OursColor}
Therm-FM-B & Pretrained & 0.046 & 0.009 & 0.094 & 0.357 & 0.034 \\
\rowcolor{OursColor}
Therm-FM-L & Pretrained & \textbf{0.031} & \textbf{0.006} & \textbf{0.069} & \textbf{0.263} & \textbf{0.023} \\

\hline
\multicolumn{7}{c}{\textit{HS-OC Refine1 ($85\times85$)}} \\
\hline
SAU-FNO & -- & 0.249 & 0.064 & 0.492 & 1.453 & 0.158 \\

\rowcolor{SotaColor}
Therm-FM-T & Scratch & 0.272 & 0.056 & 0.624 & 2.241 & 0.186 \\
\rowcolor{SotaColor}
Therm-FM-B & Scratch & 0.179 & 0.037 & 0.385 & 1.377 & 0.126 \\
\rowcolor{SotaColor}
Therm-FM-L & Scratch & 0.097 & 0.020 & 0.217 & 0.773 & 0.067 \\

\rowcolor{OursColor}
Therm-FM-T & Pretrained & 0.117 & 0.023 & 0.334 & 1.198 & 0.077 \\
\rowcolor{OursColor}
Therm-FM-B & Pretrained & 0.069 & 0.013 & 0.198 & 0.708 & 0.045 \\
\rowcolor{OursColor}
Therm-FM-L & Pretrained & \textbf{0.035} & \textbf{0.007} & \textbf{0.097} & \textbf{0.346} & \textbf{0.023} \\

\hline
\end{tabular}

\vspace{0.05in}
\footnotesize{
\textit{Notes.}
\colorbox{SotaColor}{Scratch} denotes training from random initialization.
\\
\colorbox{OursColor}{Pretrained} denotes fine-tuning from pretrained weights.
T/B/L denote Tiny/Base/Large model scales.
}
\vspace{-8pt}
\end{table}

\begin{figure*}[htbp]
    \vspace*{-4pt}
    \centering
    \includegraphics[width=\textwidth]{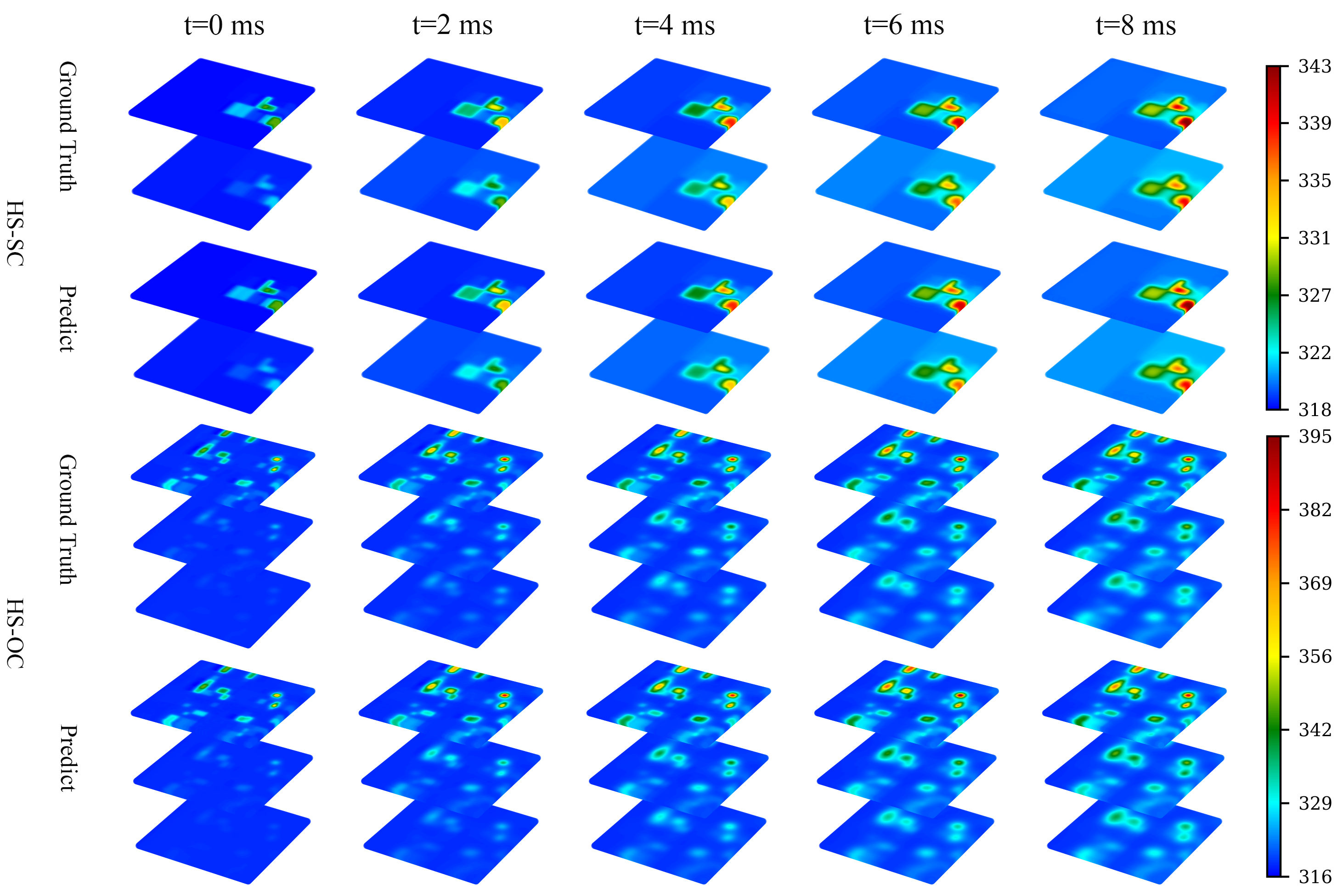}
    \caption{
    Qualitative visualization of transient thermal prediction on two representative cases, HS-SC and HS-OC. 
    Each column corresponds to a sampled time step, and each pair of rows shows the ground-truth and predicted temperature fields for one case.
    }
    \label{Fig:transient_visualization}
    \vspace*{-10pt}
\end{figure*}

\begin{figure*}[htbp]
    \vspace*{-4pt}
    \centering
    \includegraphics[width=\textwidth]{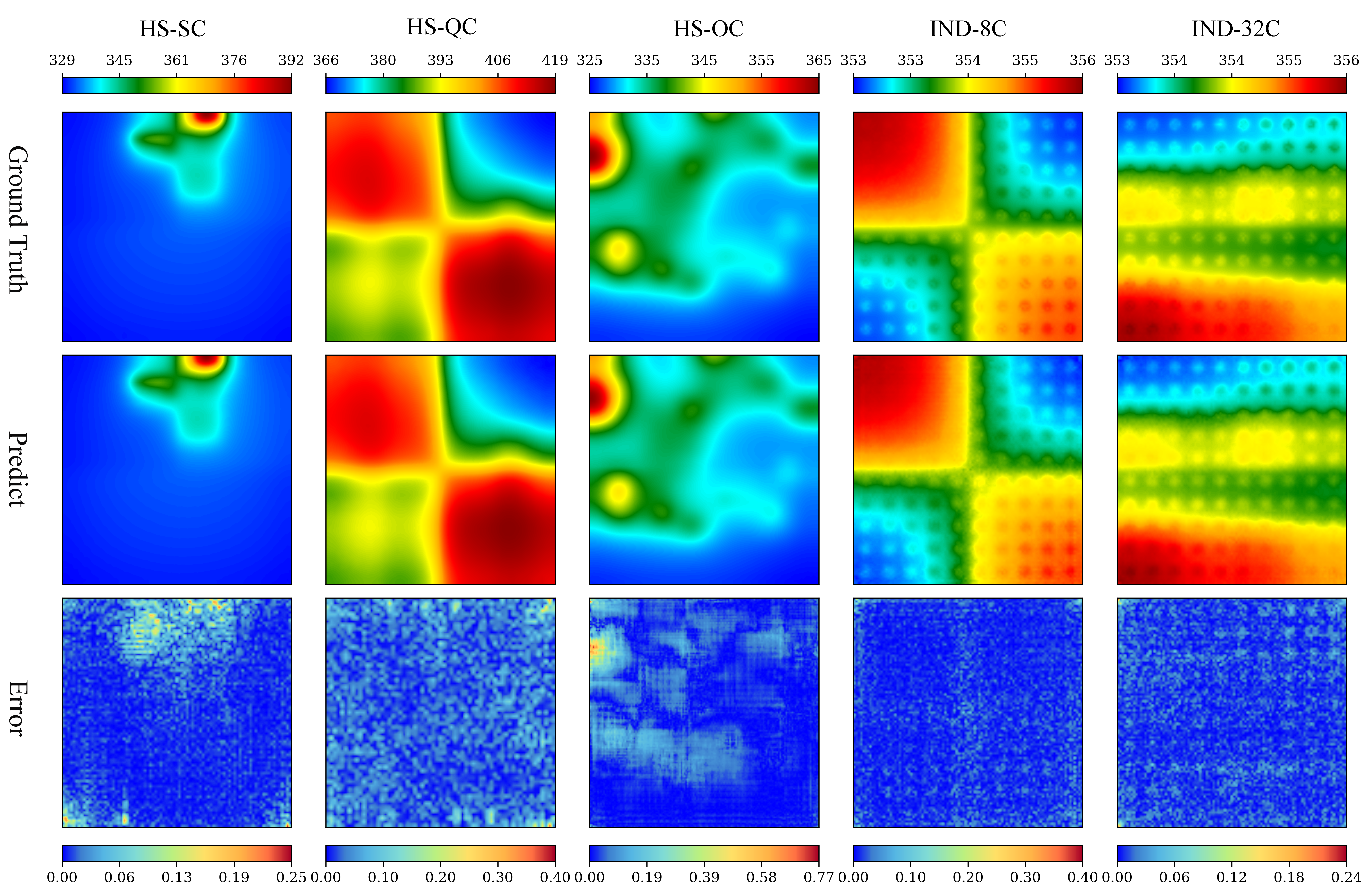}
    \caption{
    Qualitative comparison of steady-state thermal prediction results across five representative cases. 
    For each case, the first row shows the ground-truth temperature field, the second row shows the prediction of Therm-FM, and the third row shows the absolute error map.
    }
    \label{Fig:steady_visualization}
    \vspace*{-10pt}
\end{figure*}

A similar trend is observed on the more challenging HS-OC ($85\times85$) case, which involves eight cores and stronger inter-die thermal coupling than HS-QC.
Under random initialization, Therm-FM exhibits a clear scaling trend: the RMSE decreases from $0.272$ for Therm-FM-T(scratch) to $0.179$ for Therm-FM-B(scratch) and $0.097$ for Therm-FM-L(scratch).
Compared with SAU-FNO's RMSE of $0.249$, Therm-FM-B(scratch) achieves a $1.4\times$ lower RMSE, while Therm-FM-L(scratch) provides a stronger reduction of about $2.6\times$.
This indicates that the proposed architecture can benefit substantially from increased model capacity and can capture complex spatial thermal coupling even without pretrained initialization.

More importantly, pretrained initialization consistently provides substantial additional gains across all model scales.
On HS-OC, fine-tuning from pretrained weights reduces the RMSE from $0.272$ to $0.117$ for Therm-FM-T, from $0.179$ to $0.069$ for Therm-FM-B, and from $0.097$ to $0.035$ for Therm-FM-L, corresponding to about $2.3\times$, $2.6\times$, and $2.8\times$ reductions over their scratch counterparts.
Compared with SAU-FNO, the pretrained Therm-FM models achieve RMSE reductions of about $2.1\times$, $3.6\times$, and $7.1\times$ for the Tiny, Base, and Large scales, respectively.
This confirms that the performance of Therm-FM comes from both the expressive architecture and the transfer of diffusion-oriented physical priors from pretraining.
Without pretrained initialization, the model can still learn useful spatial coupling patterns as the model scale increases, but it requires more data and optimization effort to reach the same level of accuracy. 
Pretrained initialization therefore acts as a key enabler of Therm-FM's data efficiency.

\textbf{Effect of model scaling.}
We further analyze the impact of model capacity on IND-8C and IND-32C. 
As shown in Fig.~\ref{fig:param_metric_scaling}, FNO obtains only limited improvements as the parameter count increases and tends to saturate early. 
In contrast, Therm-FM shows a more stable error reduction trend across parameter scales, and its performance gap over FNO becomes larger for higher-capacity models. 
These results suggest that Therm-FM extracts greater accuracy gains from increased model capacity, demonstrating stronger scaling behavior and a higher performance ceiling.
This further supports the advantage of foundation-model-based adaptation over training from scratch for learning transferable thermal-field representations.

\textbf{Effect of multi-fidelity training.}
Table~\ref{tab:hf_align} studies the effect of multi-fidelity training on IND-32C. 
All results are evaluated against high-fidelity ground truth. 
Compared with training using only high-fidelity data, multi-fidelity training introduces a small increase in error for all evaluated methods, which is expected because low-fidelity samples may have a mild distribution gap from the high-fidelity thermal fields. 
However, this degradation remains limited, especially for Therm-FM. 
For Therm-FM-B, the Mean error only changes from $0.012$ to $0.013$, and the RMSE changes from $0.016$ to $0.018$. 
Although the Max error has a relatively larger fluctuation due to its sensitivity to local outliers, Therm-FM-B still clearly outperforms FNO and SAU-FNO under both high-fidelity-only and mixed-fidelity settings.

These results show that low-fidelity data can effectively reduce the need for expensive high-fidelity simulations while maintaining comparable prediction accuracy. 
In our setting, full high-fidelity training requires 900 high-fidelity samples, while multi-fidelity training uses the same total number of samples with a $1:3$ high-to-low fidelity ratio. 
Since generating one high-fidelity sample takes about $1766\,\mathrm{s}$ and one low-fidelity sample takes about $6\,\mathrm{s}$, the mixed-fidelity setting reduces the data-generation cost by about $74.7\%$.

\begin{table}[htbp]
\centering
\caption{Effect of multi-fidelity training on IND-32C.}
\label{tab:hf_align}
\begin{tabular}{c|c|c|c|c|c|c}
\hline
\textbf{Method} & \textbf{Mixed} & \textbf{RMSE} & \textbf{MAPE} & \textbf{PAPE} & \textbf{Max} & \textbf{Mean} \\
\hline
\multirow{2}{*}{FNO}     
        & --           & 0.165 & 0.032 & 0.317 & 1.096 & 0.126 \\
        & $\checkmark$ & 0.180 & 0.040 & 0.328 & 1.125 & 0.138 \\
\multirow{2}{*}{SAU-FNO} 
        & --           & 0.096 & 0.019 & 0.186 & 0.706 & 0.085 \\
        & $\checkmark$ & 0.109 & 0.022 & 0.191 & 0.727 & 0.095 \\
\multirow{2}{*}{Therm-FM-B} 
        & --           & 0.016 & 0.004 & 0.090 & 0.330 & 0.012 \\
        & $\checkmark$ & 0.018 & 0.004 & 0.094 & 0.336 & 0.013 \\
\hline
\end{tabular}
\vspace{2pt}
\begin{flushleft}
\footnotesize
\textbf{Note:} `--' denotes training with high-fidelity data only, while $\checkmark$ denotes multi-fidelity training with both high- and low-fidelity data.
\vspace*{-15pt}
\end{flushleft}
\end{table}

\section{Conclusion and Discussion}

This work demonstrates that the structural equivalence between heat conduction in 3D-ICs and diffusion-type PDEs is practically exploitable: pretraining on diverse PDE datasets instills physical priors that can be effectively transferred to chip-level thermal fields, enabling data-efficient adaptation without per-chip retraining from scratch.
Experiments on public HotSpot benchmarks and newly constructed industrial 3D-IC package cases confirm consistent advantages in prediction accuracy, data efficiency, cross-chip adaptation, and model scaling over existing learning-based methods trained from scratch.
The combination with a multi-fidelity training strategy that substitutes expensive high-fidelity simulations with thermal-equivalent data further reduces data generation cost without compromising prediction accuracy, making the overall framework practical for real EDA workflows.

The effectiveness of pretrained initialization warrants closer examination, particularly because simply increasing FNO's model capacity yields only marginal improvement (Fig.~\ref{fig:param_metric_scaling}), whereas Therm-FM benefits substantially from both pretraining and scaling.
We attribute this asymmetry to two factors.
First, PDE foundation models are pretrained on diverse diffusion-type problems with varying source distributions, boundary conditions, and diffusivity fields, a distribution that structurally overlaps with 3D-IC heat conduction.
This pretraining instills a diffusion-oriented physical prior: the pretrained backbone has already learned how spatial temperature gradients propagate in response to localized sources, how material interfaces perturb heat flow, and how temporal diffusion fronts evolve.
When fine-tuned on chip-level thermal data, the model refines these priors rather than constructing them from scratch, requiring far fewer target samples to reach a given accuracy level.
Second, the ablation results in Table~\ref{tab:ablation_pretrain} show that pretrained initialization is critical under limited target data.
Thermal predictors trained from scratch tend to fit design-specific power--temperature mappings and generalize poorly to unseen power distributions.
By contrast, the pretrained backbone provides transferable diffusion-oriented priors, leading to more stable adaptation and lower sensitivity to training set size.
This distinction has a practical implication: for new chip designs with scarce high-fidelity data, foundation-model adaptation can approach or surpass the full-data performance of methods trained from scratch using far fewer target samples, making it more suitable for limited-data thermal modeling.

The thermal-equivalent model provides a low-cost approximation for multi-fidelity training, achieving over $300\times$ simulation speedup with a marginal steady-state error of $\sim 1\%$, even for dense TSV configurations with volume fractions as high as $20\%$. 
In Therm-FM, these low-fidelity data are used only in the first stage to provide coarse thermal-domain adaptation, while the second-stage high-fidelity FEM samples correct the residual bias introduced by the homogenization of TSV and $\mu$bump structures.
As shown in Table~\ref{tab:hf_align}, this strategy achieves accuracy close to high-fidelity-only training, indicating that the low-fidelity data improve adaptation without dominating the final prediction.
For extremely high TSV densities, where EMT approximation error increases, additional high-fidelity calibration may be required.

Therm-FM still requires a small number of target-domain samples for reliable adaptation, and its generalization is bounded by the physical regimes and geometric structures covered in the pretraining data.
Direct inference-time comparison with commercial FEM solvers is not a focus of this work; for trained surrogate models, orders-of-magnitude inference speedup over FEM is generally expected, and the more critical bottleneck in practice is data generation cost and cross-design adaptation, which this work directly addresses.
In EDA applications, high-fidelity thermal data remain expensive to generate and publicly available datasets are scarce, which will likely constrain further scaling of thermal foundation models.
Future work will explore larger and more diverse pretraining datasets, broader industrial thermal benchmarks, and multi-physics extensions to further advance foundation-model-based simulation for EDA. To facilitate reproducible research and future development, we release all datasets, code, and pretrained models at \url{https://github.com/haiyangxin/Therm-FM}.

\section*{Acknowledgment}
The authors acknowledge the use of generative AI tools, including GPT-based language models, solely for language polishing, grammar checking, and improving the readability of early manuscript drafts. All technical ideas, methodologies, experiments, data analyses, and conclusions were developed, verified, and approved by the authors.

\bibliographystyle{IEEEtran}
\bibliography{references}

\newpage

\vfill

\end{document}